\documentclass[12pt,draftcls,onecolumn]{IEEEtran} 		

\usepackage{amsmath,amssymb,amsfonts,xpatch}
\usepackage{cite}
\usepackage[ruled,vlined]{algorithm2e}
\usepackage{setspace}

\usepackage{graphicx}
\usepackage{cool}
\usepackage{caption}
\usepackage{subfigure}
\usepackage{rotating}
\usepackage{graphics}
\usepackage{nccmath}
\usepackage{pdfpages}
\usepackage{mathcomSTEv4}
\usepackage[printonlyused]{acronym}

\usepackage[margin=2cm]{geometry}
\usepackage{adjustbox}
\usepackage{multirow}

\usepackage{hyperref}
\hypersetup{colorlinks,citecolor=green}

\usepackage{tikz,pgfplots}
\pgfplotsset{width=10cm,compat=1.9}
\usetikzlibrary{calc}

\usepackage{pgfplotstable}
\usetikzlibrary{positioning}
\usetikzlibrary{pgfplots.groupplots}
\usetikzlibrary{arrows}

\usepackage[printonlyused]{acronym}

\acrodef{mac}[MAC]{\emph{multiple access channel}}
\acrodef{oimac}[OIMAC]{\emph{optical intensity multiple access channel}}
\acrodef{snr}[SNR]{\emph{signal-to-noise ratio}}
\acrodef{awgn}[AWGN]{\emph{additive white Gaussian noise} }
\acrodef{mine}[MINE]{\emph{mutual information neural estimator}  }
\acrodef{dine}[DINE]{\emph{directed information neural estimator}} 
\acrodef{smile}[SMILE]{\emph{smoothed mutual information lower-bound estimator}  }

\acrodef{nmie}[NMIE]{\emph{\emph{neural mutual information estimation}} }
\acrodef{nit}[NIT]{\emph{neural distribution transform} }
\acrodef{ndt}[NDT]{\emph{neural distribution transform} }
\acrodef{dv}[DV]{\emph{Donsker-Varadhan}  }
\acrodef{cmi}[CMI]{\emph{conditional mutual information}  }
\acrodef{mi}[MI]{\emph{mutual information}  }

\newtheorem{theorem}{Theorem}

\newtheorem{rem}{Remark}

\newcommand{\clip}{\mathrm{clip}}

\newcommand{\nats}{\mathrm{nats}}
\newcommand{\db}{{\rm dB}}

\definecolor{orange}{RGB}{255,127,0}
\definecolor{byzantium}{rgb}{0.44, 0.16, 0.39}
\definecolor{byzantine}{rgb}{0.74, 0.2, 0.64}

\newcommand{\ste}{ \color{blue}}
\newcommand{\fm}{ \color{blue}}
\newcommand{\nf}{ \color{blue}}

\title{
A Perspective on Neural Capacity Estimation: Viability \& Reliability
}

\author{\IEEEauthorblockN{Farhad~Mirkarimi}, 
	\and
	\IEEEauthorblockN{Stefano~Rini},
 	\and
 	\IEEEauthorblockN{Nariman~Farsad}
}

\begin{document}

\maketitle

\begin{abstract}
%

 Recently, several methods have been proposed for estimating the mutual information from sample data using deep neural networks.
    %
    These estimators ar referred to as
    \ac{nmie}s.
    {\ste \ac{nmie}s differ from other approaches as they are data-driven estimators. 
    As such, they have the potential to perform well on a large class of capacity problems. 
    %
    %
    %
    In order to test the performance across various  \ac{nmie}s, it is desirable to establish a benchmark  encompassing the different challenges of capacity estimation. 
    This is the objective of this paper. 
    In particular, we consider three scenarios for benchmarking: (i) the classic AWGN channel,  (ii) channels continuous inputs -- 
    optical intensity and peak-power constrained AWGN channel  (iii) channels with a discrete output-- i.e., Poisson channel.
    We also consider the extension to the multi-terminal case with (iv) the AWGN and optical MAC models.
    We argue that  benchmarking a certain \ac{nmie} across these four scenarios 
    provides a substantive test of performance.
    %
}
%
In this paper we study the performance of  \ac{mine}, 
\ac{smile}, 
and  \ac{dine}. 
and provide insights on the performance of other methods as well. 
%
%
%
%
%
\ste{
To summarize our benchmarking results, MINE provides the most reliable performance.
%
}
\end{abstract}


\section{Introduction}

Determining the capacity of a channel and the optimal input distribution are problems of fundamental importance in communication scenarios of practical relevance. 
%
The value of capacity and the optimal input distribution provide critical insight on the choice of coding rates and input constellation shape, respectively.
{\ste
The traditionally, the literature on capacity computation has straddled two tendencies: that of (i) considering channel models that have an elegant formulation  for which capacity is evaluated in closed-form through mathematical manipulations, versus that of   
(ii) considering channel models obtained from a practical communication setting for which capacity is determined through numerical optimization techniques.
Neural mutual information estimators (\ac{nmie})  represent a novel approach to the analysis and design of communication systems that does not incur in the above limitations, since  (i) they do not require the human mathematical modelling of the channel behaviours, as well as (ii) can efficiently tackle the inherent complexity of the capacity estimation problem. 
Instead, \ac{nmie}S only require the ability to simulate the channel input/output relationship and use coupled DNN to estimate the capacity of the model:
%
(i) one DNN  that generates samples from the estimated optimal input distribution that, and (ii) a DNN that estimate the mutual information (MI) between the channel input and the channel output. 
In this sense, \ac{nmie}s can be consider as ``universal'' estimators in the sense that they only require data samples from the input/output relationship.

Since \ac{nmie}s are numerical methods, any question regarding their performance can only be investigated by benchmarking. 
A natural question that then arises  is then

\begin{quote}
What constitutes a substantive and yet minimal set of channels to test the performance of an \ac{nmie}?
\end{quote}
In this paper we make a first attempt in defining such a benchmark for \ac{nmie}. 
More specifically, We consider the following models: AWGN channel, optical intensity,  peak-power constrained AWGN, and the Poisson channel. 
For each channel, we extensively test the performance of three \ac{nmie}s over different SNR ranges: \ac{mine}, \cite{belghazi2018mine},  
\ac{smile}, \cite{smile2020},  and  \ac{dine}. \cite{aharoni20_CapMemChan}. 
We give quantitative as well as qualitative assessment of their performance in the various regimes, as well as provide insight on the training process. 
In particular, we test the ability of these \ac{nmie}s to estimate optimal input distributions with large support, with discrete mass points, and with mass points that grow large as the SNR increases. 

We argue that this rigorous benchmarking is necessary not only to discover the relative advantages of the various approaches, but also for the development of new \ac{nmie}s which are able to fill the performance gaps of existing solutions. 
}

\subsection{Relevant Literature}

The computation of capacity using  numerical methods most  dates back to the well-known works of Blahut and Arimoto \cite{1054855, 1054753}. This algorithm consists of an iterative alternating maximization method using the primal and dual formulation of the capacity problem \cite[Sec. 4]{gallager1968information}.
Various authors have considered extension and refinements to this original algorithm these include \cite{caocapacity11}\cite{perm}\cite{Far20DAB}.

Similarly, computing closed-form expressions for the capacity region of general multi-user channels, except in few cases, has proved to be challenging. The capacity region of a two and three user memoryless \ac{mac} was formulated in \cite{ahl1973multi}. Later, \cite{msalehi} considered the capacity region of \ac{mac} with correlated sources and derived its achievable rates region.  In \cite{wolf}, it was shown that feedback can increase the rate region of memoryless \ac{mac}, while \cite{permuter2009capacity} derived the capacity region of finite state \ac{mac}s and provided multi-letter expressions for the capacity region.
Finally, a number of other works including \cite{oimac1, oimac2} consider the capacity of optical intensity \ac{mac} with both average and peak power constraints on the channel input, and derive different inner and outer bounds for the capacity region. Despite all these results, numerical computation of the capacity region of the arbitrary memoryless \ac{mac} has proved to be challenging \cite{calvo10_BAMAC}.

Recently deep learning has become a new and powerful tool in communications with possible applications for designing new channel codes, new modulation schemes, and evaluating achievable information rates \cite{9085350}. Reinforcement learning is used to evaluate the capacity of point-to-point channels with feedback in \cite{aharonicomputing}. Recent advancements in estimating mutual information from data samples using deep learning methods \cite{belghazi2018mine} has resulted in several new works. In \cite{yechannel}, the estimator in \cite{belghazi2018mine} is used to produce efficient joint encoder and decoders for modulation (with a low probability of error) by maximizing mutual information between inputs and outputs of the channel. In \cite{aharoni20_CapMemChan}, a capacity estimation algorithm is developed for continuous channels with feedback using a deep learning-based estimator of directed information. Finally in \cite{fm0}, a neural capacity estimation approach is proposed for the MAC using $\chi ^2$ bound.

{\ste 

From a more general perspective, in recent years, end-to-end learning using machine learning has been proposed for system design in communication systems \cite{jakob1, ye1, ye2}. 
Since the ultimate goal of communication system design is to increase reliable information transmission rate, \ac{mi} has been proposed as a loss for optimizing system design in \cite{stark1, wunder1, wunder2, capauto}.

}

 \subsection{Contributions}

{\ste

In this work, we numerically investigate the performance of various \ac{nmie} over a relevant set of channels and for a relevant set of parameter regimes. %

%
We focus on two capacity problems: (i) problem of finding the capacity of memoryless point-to-point channels with continuous alphabets and (ii) the problem of finding the capacity region of the memoryless MAC. 
}

{
\ste
Generally speaking, \ac{nmie} obtain a estimate of the capacity value by iteratively repeating two steps:  first, (i) the mutual information between the channel input and the channel is estimated as a function of the input distribution through a deep neural network (DNN), then  (ii) this mutual information is maximized as a function of the input distribution, which is also expressed through a DNN. 
More details are presented in Sec. \ref{sec:Neural Capacity Estimator}.
}

%
{
\ste
For this class of capacity estimators, we benchmark performance over  (i) the classic AWGN channel, (ii) channels continuous inputs – optical intensity
and peak-power constrained AWGN channel (iii) channels with a discrete output– i.e., Poisson channel.
We also consider the extension to the multi terminal case with (iv) the AWGN and optical MAC models. 
%

}

The reminder of the paper is organized as follows.

\medskip
\noindent
$\bullet$ {\bf Sec. \ref{sec:NeuralMIestimator}: }
We start the paper by introducing the different \ac{nmie} proposed in the literature. Generally speaking, there are two general classes of estimators.
A first class of estimator performs  a direct estimation of the mutual information: among such estimators we study \ac{mine} \cite{belghazi2018mine} and \ac{smile} \cite{smile2020}.
Another class of estimators estimates the entropy  based on reference random variables\cite{chan2019neural}:
among such estimators we consider \ac{dine} \cite{aharoni20_CapMemChan} and 
InfoNCE \cite{oord2018representation}.

Note that these entropy-based methods can be used for estimation of other information theoretic measures besides mutual information such as directed information and conditional mutual information.
This approach was explored in \cite{aharoni20_CapMemChan} and \cite{fm0}.

\medskip
\noindent
$\bullet$ {\bf Sec. \ref{sec:Neural Capacity Estimator}:}
We discuss how the \ac{nmie} presented in Sec. \ref{sec:NeuralMIestimator} are employed for channel capacity estimation. 
This can be accomplished through an alternating optimization:
first (i) a  \ac{ndt}
network trained to produce an  estimate of the capacity-approaching distribution;
successively (ii) the \ac{nmie} is trained to find a tight bound on the mutual information between channel input and output. 
By alternating these two optimization steps, convergence to a tight estimation of both the capacity and the capacity achieving distribution can be obtained. 
In this section, we shall also discuss the role of the initialization of both the \ac{ndt} and the \ac{nmie} is discussed.



\medskip
\noindent
$\bullet$ {\bf Sec.  \ref{sec:Channel Models and Related Results}: }
In this section, we introduce the channel models that we select for benchamrking,
that is: the  1) AWGN channel with an average power constraint, 2) the optical intensity channel, 3) the AWGN with a peak power constraint, and 4) the Poisson channel. We also consider the extension of the proposed approach to the multiple access channel (MAC) by considering 1) the AWGN MAC and the 2) the optical intensity MAC. We present different capacity results in each case.
%

\medskip
\noindent
$\bullet${ \bf Sec. \ref{sec:Comparison existing results}:} 
 We begin our benchmarking study by testing the performance of the various \ac{nmie} in Sec. \ref{sec:NeuralMIestimator}  for the  case of the  classic AWGN channel. 
%
%
Since the capacity and the optimal input distribution for this classic channel are well-known, we can perform a detailed comparison of different methods.  
%
%
Various insights on the precision, numerical stability, and computational complexity of the different algorithms are provided. 
Overall, \ac{mine} emerges as a reliably and efficient \ac{nmie} estimator for capacity estimation. 

\medskip
\noindent
$\bullet${ \bf Sec. \ref{sec:Numerical Study}:}
The insights gained on the performance of the various \ac{nmie} estimators on AWGN channel are extended to estimating the capacity of the other channels presented in Sec.  \ref{sec:Channel Models and Related Results}. 
Based on our numerical evaluations we observe that for the point-to-point channels, \ac{nmie} based on direct estimation of the mutual information give more accurate estimates of the capacity and estimated the optimal input distribution more closely. 
We also show that neural capacity estimation can give tighter bounds for channels with only upper and lower bounds on capacity such as the optical intensity channel.
In the latter part of the section, in Sec. \ref{sec:Multiple Access Channel (MAC)}, we consider the problem of estimating the capacity region of the Gaussian and optical MAC channel.

\bigskip

Before proceeding further, let us introduce the notation employed in the paper.

 \subsubsection*{Notation}
The sets $\{i, i+1, \ldots, j\} \subseteq \Nbb$  and $\{1, \ldots, j\}$ are abbreviated as $[i:j]$ and $[j]$, respectively.	
The $L$-norm of the vector $\xv \in \Rbb^d$ is indicated as $\|\xv\|_L$.
Random variables (RVs) are indicated with capital roman letters, i.e. $X$,  their support with the same calligraphic letter, i.e. $\Xcal$, and  their pdf as  $P_X(x)$ or $Q_X(x)$, and when the context is clear as $P(x)$ or $Q(x)$. 
The normal distribution is indicated as $\Ncal$, the tail probability function for the standard normal distribution in $x$ is indicated  $\Qcal(x)$.
%
In the remainder of the paper, we consider four measures with respect to RVs: the Kullback–Leibler (KL) divergence, entropy, mutual information (MI), and the $\chi^{2}$ divergence, respectively defined as
\begin{flalign}
D(P || Q) & = \int_{}^{} P(x) \log \frac {P(x)}{Q(x)}   d x, \label{eq:KL}\\
 h(X) & =\mathbb{E}_P [-\log(Q(x))]-D(P||Q) \label{eq:entropy}\\
I(X,Y) & = D(P(x,y) || P(x) P(y))\label{eq:MI}\\
\chi^{2}(P||Q) & =\int_{}^{}\frac{(P(x)-Q(x))^{2}}{Q(x)} d x,\label{eq:chi}
\end{flalign}
respectively, where we have agreed that $ 0 \log 0 / 0=0$.
The binary entropy is defined as $H_2(p)=p \log(p)+(1-p)\log(1-p)$.

The set of RV $\{\Xv_m,\ldots \Xv_n\}$ is indicated as $\{\Xv_i\}_{i \in [N:M]}$ .
For brevity, we define $ \{\Xv_i\}_{i=-\infty}^\infty=\{\Xv_i\}$.  
When this set of random variables is used to construct a random vector, we employ the notation  $\Xv_m^n = [X_m,X_{m+1},\ldots, X_n]$ with $n \ge m$.  
Again, when $m=1$, the subscript is omitted, i.e. $\Xv_1^n=\Xv^n$. 
Given the scalar $P$, the value in dB is indicated as $P_{db}$, i.e. $P_{db}=10 \log_{10}(P)$.

\medskip

The code used in the remainder of the paper can be found online here:
https://github.com/Farhad-Mrkm/Benchmarking-Neural-Capacity-Estimators

  

\section{Neural Estimation of Mutual information}
\label{sec:NeuralMIestimator}

In recent years, there have been many different methods proposed for estimating mutual information using neural networks \cite{belghazi2018mine, 2019variational, chan2019neural, smile2020, fm0}.
We refer to these approaches as \ac{nmie}.
We review these methods in this section and then in the next section, we discuss how they can be employed to estimate capacity and the optimal input distribution.

The \ac{nmie} methods can be divided into two groups:  \ac{nmie} that evaluate the mutual information in \eqref{eq:MI} directly by using bounds on KL divergence \cite{belghazi2018mine, 2019variational, smile2020}, and  (ii) \ac{nmie}  that break the mutual information into entropy terms and estimate each entropy term separately using bounds on KL divergence \cite{chan2019neural, aharoni20_CapMemChan, fm0}. 
We start this section by presenting the bounds on KL divergence, which are used by these \acp{nmie}.

\subsection{Bounds on Information-theoretic Measures}
\label{subsec:KLbounds}

In this section, we provide lower and upper bounds on KL-divergence. Then in the next subsections, we use these bounds to describe the various methods that have been proposed in the literature for estimating mutual information using neural networks.    

The first bound considered is the Donsker-Varadhan~\cite{donsker1983} bound rewritten in the theorem below for convenience.
\begin{theorem}[Donsker-Varadhan representation \cite{donsker1983}]
	\label{thm:KLlower}
Consider the RV $X$ and two probability measures over $\Xcal$, $P$ and $Q$, then the KL-divergence can be rewritten as
	\begin{equation}
	\label{eq:dv_bound}
			D(P || Q) = \sup_{T \in \mathcal{T}} \mathbb{E}_{P}[T(X)] - \log \lb \mathbb{E}_Q\left[e^{T(X)}\right] \rb,
	\end{equation}
	where $\mathcal{T}$ is the set of all functions with finite expectations in~\eqref{eq:dv_bound}.
\end{theorem}

Another lower bound for KL-divergence is obtained in \cite{2019variational, fm0}, and is presented here for convenience.

\begin{theorem}[Tractable Unnormalized Barber and Agakov (TUBA) lower
bound \cite{2019variational, fm0}]
\label{th:tuba}
Consider the RV $X$ and two probability measures over $\Xcal$, $P$ and $Q$, then the KL-divergence is lower-bounded as
\begin{equation}
    \label{eq:tubalower}
    D(P || Q) \geq \mathbb{E}_{P}[T(X)]-\lsb \frac{\mathbb{E}_Q\left[e^{T(X)}\right]}{\alpha}+\log(\alpha)-1\rsb,
\end{equation}
where $\alpha$ is a positive constant.
\end{theorem}

An upper bound on KL-divergence using $\chi^2$-divergence can be obtained using \cite[Theorem 23]{verdu2016fDiverg}, which is rewritten in the theorem below for convenience.
 \begin{theorem}[KL divergence upper bound \cite{verdu2016fDiverg}] 
	\label{thm:KLupper}
	Consider a RV $X$ and two probability measures over $\mathcal{X}$, $P$ and $Q$.
 	The KL divergence is then upper bounded as
\begin{flalign}
	D(P||Q) & \leq \log(1+\chi^{2}(P||Q))  \\
	& \quad \quad -\frac{1.5\,(\chi^{2}(P||Q))^{2}\log e}{(1+\chi^{2}(Q||P))(1+\chi^2(P||Q))^{2}-1}
	\nonumber 
	\end{flalign}
\end{theorem}
{\fm In order to make this bound amenable to computation using data driven methods we use a variational form for $\chi^2$ from\cite{yuri}:
\begin{equation}\label{mainchi}
    \chi^2(P||Q)=\sup_{T \in \mathcal{T}} \mathbb{E}_{P}[T(X)] - \log \lb \mathbb{E}_Q\left[{T^{2}(X)}\right] \rb-1
\end{equation}}
{ \fm As a result we have following variational form for the upper bound in \eqref{mainchi}
\begin{flalign}
	D(P||Q) & \leq \sup_{T \in \mathcal{T} }\log(\mathbb{E}_{P}[T(X)] - \log \lb \mathbb{E}_Q\left[{T^{2}(X)}\right] \rb)  \\
	& \quad \quad -\frac{\sup_{T \in \mathcal{T} }1.5\,(\mathbb{E}_{P}[T(X)] - \log \lb \mathbb{E}_Q\left[{T^{2}(X)}\right] \rb-1)^{2}\log e}{(\sup_{T \in \mathcal{T} }\mathbb{E}_{Q}[T(X)] - \log \lb \mathbb{E}_P\left[{T^{2}(X)}\right] \rb)(\sup_{T \in \mathcal{T} }\mathbb{E}_{P}[T(X)] - \log \lb \mathbb{E}_Q\left[{T^{2}(X)}\right] \rb)^{2}-1} \nonumber \\
	& \triangleq \chi^{2}_{\mathsf{UP}}(P||Q).
	\label{eq:X up}
\end{flalign}}

\textcolor{blue}{Where we have used monocity of $\log$}.
 \textcolor{blue}{In next section we use this bound to find our proposed explicit variational lower bound on mutual information, and we prove its consistency.}
The next section, Sec. \ref{sec:directMIestimator},  is devoted to present different neural mutual estimation approaches that utilize the results in Thm. \ref{thm:KLlower}-\ref{thm:KLupper}.

\subsection{Direct Mutual Information Estimation}
\label{sec:directMIestimator}

In this section, we review the \acp{nmie} that estimate the mutual information in \eqref{eq:MI} through bounds on the KL divergence term on the right hand side of \eqref{eq:MI}.

\subsubsection{MINE}
One of the earliest attempt to NMIE estimation can be found in \cite{belghazi2018mine}, where the 
\ac{mine} is proposed. 
The MINE is obtained from the the \ac{dv} representation in Th. \ref{thm:KLlower} when a DNN is used to represent $T(X,Y)$ in \eqref{eq:dv_bound}. Note that here the KL divergence is between the joint distribution and the product of the marginals as in \eqref{eq:MI}, and hence the function $T(.)$ is a function of $X$ and $Y$.
It is noted in \cite{belghazi2018mine} that the optimization of $T(X,Y)$ using a DNN  through stochastic gradient descent (SGD) is generally challenging. 
This is because using a naive gradient estimate over the samples of the mini-batch leads to a biased estimate of the full gradient. Exponential moving average is proposed for mitigating this problem. However, this leads to an MI estimator with large variance, as noted in \cite{2019variational}.
In  \cite{belghazi2018mine},  the MINE is shown to be effective in preventing the mode collapse in generative adversarial networks (GANs).

\subsubsection{SMILE}
To address the large variance of the MINE, the authors of \cite{smile2020} propose the \ac{smile}. In this method, the \ac{dv} representation in Theorem \ref{thm:KLlower} is rewritten as in the following theorem. 
\begin{theorem}[SMILE \cite{smile2020}]
Consider the RV $X$ and two probability measures over $\mathcal{X}$, $P$ and $Q$. The KL-divergence can then be approximated as
\begin{flalign}
	D(P || Q) \approx \sup_{T \in \mathcal{T}} \mathbb{E}_{P}[T(X)] - \log  \mathbb{E}_Q\big[\clip(e^{T(X)},\tau)\big],
    \label{eq:smile bound}
\end{flalign}
where $\clip(x,\tau) \triangleq \min\{\max\{x,e^{-\tau\}},e^\tau\}$.
\end{theorem}
SMILE leverages the bound in \eqref{eq:smile bound} as the clipping function $\clip$  is equivalent to clipping the log density ratio estimator in the interval $[-\tau,+\tau]$. Similar to \ac{mine} mutual information can be estimated by using a neural network to represent $T(.)$, and optimizing the neural network using gradient descent.
The choice of $\tau$ crucially affects the bias-variance trade-off. With smaller $\tau$, the variance of the estimator is reduced at the cost of potential increase to its bias. 

\subsubsection{Reverse Jensen Inequality}
Another approach for addressing the high variance in estimation of the mutual information caused by the second expectation in the DV representation uses a reverse Jensen inequality \cite{wunder2021reverse}. In this case, the mutual information is lower-bounded as
\begin{align*}
    I(X; Y) \ge \sup_{T \in \mathcal{T}} \mathbb{E}_{P}[T(X)] - \frac{1}{\zeta_b(a)} \mathbb{E}_Q \left[ \log \left( \frac{1+e^{T(X)}}{c} \right) \right],
\end{align*}
where $\zeta_b(a)$ is defined in \cite[Lemma~1]{wunder2021reverse}, and $c$ is a scalar parameter. One of the challenges of this approach is that the parameters on the right hand side must be tuned. Another is that the proposed reverse Jensen inequality in \cite{wunder2021reverse} may not be tight for all settings.  \subsubsection{InfoNCE Contrastive Lower Bound\cite{oord2018representation}}
This bound provides a method for computing MI directly using $K$ samples of joint  distribution of $X,Y$. This bound is given by:
$$I(X; Y) \ge \mathbb{E}_{p^K(x,y)}\left[\frac{1}{K} \sum_{i=1}^K \log \frac{e^{\color{black}T_\phi(x_i, y_i)}}{\frac{1}{K}\sum_{j=1}^K e^{\color{black} T_\phi(x_j, y_i)}}\right],$$
where $p^K(x,y)$ is the distribution of $K$ independent samples from the joint distribution: $\prod_{j}^{} p(x_j,y_j)$.

\subsubsection{Variants of Donsker-Varadhan \cite{sreekumar2022neural}}
{\fm This bound has been extensively used in the literature of $f$-divergence estimation \cite{sreekumar2022neural}, and is used as loss function for generative adversarial networks \cite{nowozin2016f}. Its representation is based on the Legendre-Fenchel duality and is expressed as:\begin{equation}
    	D(P || Q) = \sup_{T \in \mathcal{T}} \mathbb{E}_{P}[T(X)] -\lb \mathbb{E}_Q\left[e^{T(X)}-1\right] \rb.
\end{equation}
}
\subsection{Entropy-based Mutual Information Estimation}

\noindent Another technique for estimating MI, rewrites MI in terms of a number of entropy terms and uses the variational bounds on KL divergence to estimate the entropy terms from data.  
Specifically, using \eqref{eq:entropy}, and known arbitrary i.i.d. reference random variables $X^{'}$ and $Y^{'}$, the mutual information can be expressed as
{
{\fm 
	\begin{align}
		I(X;Y) &= h(X) + h(Y) - h(X,Y),  \nonumber \\
		& =D(P_{X,Y}||Q_{X^{'},Y^{'}})-D(P_X||Q_{X^{'}})-D(P_Y||Q_{Y^{'}}). \label{eq:decompose}  
	\end{align}}
}%
Note that since we can choose the reference distributions to be i.i.d., the cross-entropy terms in~\eqref{eq:entropy} will cancel out leaving only the KL-divergence terms. We now present the methods that use this approach for estimating mutual information.

\subsubsection{DINE}
The \ac{dine} method estimates the directed information, which is used to compute the capacity of channels with feedback \cite{aharoni20_CapMemChan}. It can also be used to estimate the MI and the capacity of memoryless channels.  Instead of using \eqref{eq:decompose} to decompose mutual information, \ac{dine} uses $h(Y)$ and $h(Y|X)$ and hence has two KL-divergence terms to estimate instead of the three in \eqref{eq:decompose}. Each KL-divergence term is then computed using \eqref{eq:dv_bound} adopted from~\cite{chan2019neural}. Since the KL divergence term corresponding to $h(Y|X)$ is positive while the KL divergence term corresponding to $h(Y)$ is negative, and \eqref{eq:dv_bound} is a maximization, the estimated MI is neither a lower nor an upper bound. To support feedback and channels with memory, \ac{dine} uses a recurrent neural network to estimate the directed information and hence it is slower to train for memoryless channels without feedback. This can be easily resolved however by using fully connected layers to represent the function $T$ as is done in \cite{fm0}. %

\subsection{Chi-Square Bound}
\textcolor{blue}{ 
 In this subsection, We use the upper bound on the chi-square ($\chi^2$) divergence to find an explicit lower bound on MI.
A variant of this method was employed in \cite{fm0} to compute the boundaries of the capacity region of the multiple access channels. This method uses the decomposition in \eqref{eq:decompose}. 
For the positive KL divergence terms in \eqref{eq:decompose},  \eqref{eq:tubalower} 
is used where $T$ is represented and optimized using a neural network. For the negative KL divergence terms in \eqref{eq:decompose}, the upper bound $\chi^{2}_{\mathsf{UP}}(P||Q)$ from Theorem \ref{thm:chi1} is used, where these terms are replaced by a variational form of $\chi^2$ divergence. This is replaced the  histogram based plug-in estimator which initially was used in \cite{fm0}, and allows for flow of gradient in the upper bound as well. \textcolor{blue}{A detailed discussion about consistency of this estimator is provided in the appendix.}} 
The following theorem summarizes this approach:
\begin{theorem}[ $\chi^2$ Bound] 
\label{thm:chi1}
The MI between the RVs $X$ and $Y$ is lower bounded as
\begin{flalign}
    I(X;Y)\geq & \sup_{T \in \mathcal{T}}\mathbb{E}_{P}[T(X,Y)] - \log[\mathbb{E}_{Q}[e^{T^(X^\prime,Y^\prime)}]]\nonumber \\
 & \qquad -\chi^{2}_{\mathsf{UP}}(X||X^{'})-\chi^{2}_{\mathsf{UP}}(Y||Y^{'}), \label{eq:chisqrbound} 
\end{flalign}
and for $\chi^{2}_{\mathsf{UP}}$ in \eqref{eq:X up}.
\end{theorem}
We refer to this method as $\chi^{2}$ bound in the paper. 
\subsection{Extension to Conditional Mutual Information }

In this subsection we extend the \ac{nmie} methods to the estimation of the \ac{cmi}, which is needed for estimating the capacity region of the \ac{mac}. Just like the MI estimation, the \ac{cmi} can be estimated using several different approaches.

\subsubsection{Direct Conditional Mutual Information Estimation}
The \ac{cmi} can be evaluated using the \ac{dv} bound in Theorem \ref{thm:KLlower} directly using
\begin{align}
\label{eq:cmi_def}
I(X;Y \mid Z) = D(P_{X,Y,Z} || P_{X, Z} P_{Y|Z}).
\end{align}
The main challenge in using the \ac{dv} bound for estimating \ac{cmi} from data samples is creating training samples from $Q \equiv P_{X, Z} P_{Y|Z}$. In estimating MI, the samples for the product of the marginals are easily obtained using random permutations on the samples of $x$ and $y$ drawn i.i.d. from the joint distribution, i.e., $(x,y) \sim P_{X,Y}$. Several methods have been considered for drawing samples from $P_{X, Z} P_{Y|Z}$, given a data from the joint distribution using conditional generative adversarial network (CGAN), conditional variational autoencoder (CVAE), and nearest neighbors sampling \cite{mukherjee20_CCMI, mol21_neural_cmi}.

\subsubsection{Mutual Information Based Conditional Mutual Information Estimation}
Another approach for estimating \ac{cmi} relies on the the decomposition of the \ac{cmi} into difference of MI terms as:
\begin{flalign}
\label{eq:cmi_decomp}
I(X; Y \mid Z) = I(X;Y,Z) - I(X;Z).
\end{flalign}
Each MI term in \eqref{eq:cmi_decomp} could be estimated using one of the \ac{nmie} methods such as MINE and SMILE method. Note that since the consistency of these estimators has been proved in \cite{belghazi2018mine,smile2020} for MI estimation, the \ac{cmi} estimation using this approach will also be consistent.

\subsubsection{Entropy Based Conditional Mutual Information Estimation}
Finally, \ac{cmi} can be decomposed into a summation of entropy terms, which can then be evaluated by using arbitrary reference random variables $X^{'}$, $Y^{'}$, and $Z^{'}$. Specifically, \ac{cmi} can be decomposed as:
\begin{align}
I(X;Z\mid Y) = & h(Y,Z)+h(X,Y)-h(Y)-h(Y,X,Z) \nonumber \\
=&D(P_{X,Y,Z}||Q_{X^{'},Y^{'},Z^{'}})+D(P_{Y}||Q_{Y^{'}}) \nonumber \\
&\quad -D(P_{X,Y}||Q_{X^{'},Y^{'}})-D(P_{Y,Z}||Q_{Y^{'},Z^{'}}), \label{eq:condMIKL}
\end{align}
where the last equality follows from the independence of $X^{'}$, $Y^{'}$, and $Z^{'}$, which cancels out the cross-entropy terms from \eqref{eq:entropy}. The four KL-divergence terms can then be estimated using any of the prior methods based on the \ac{dv} bound such as \ac{mine} or \ac{smile}.

 \section{Neural Capacity Estimation}
\label{sec:Neural Capacity Estimator}

\begin{figure}
    \centering
    \includegraphics[width=\linewidth]{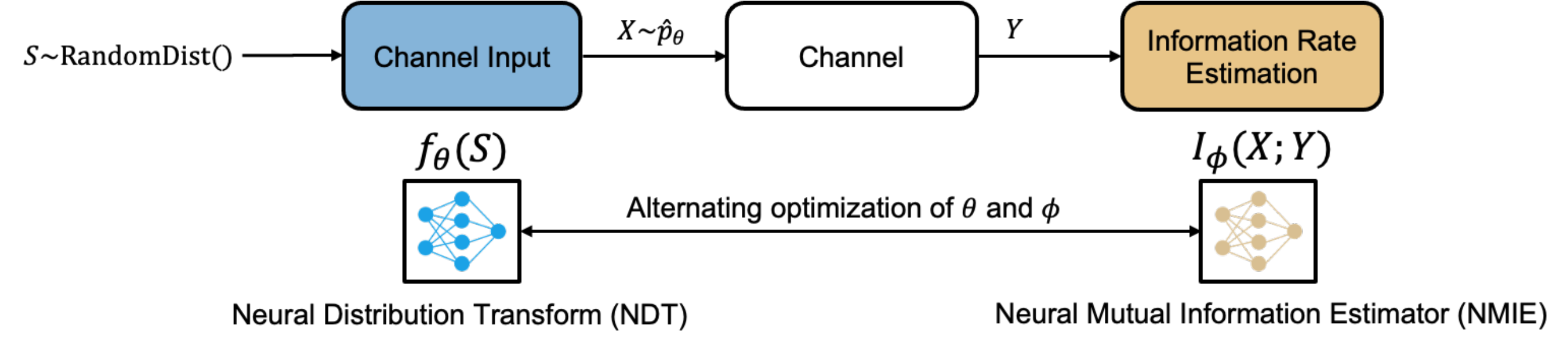}
    \caption{The block diagram of the neural capacity estimator.}\label{fig:nceBlock}
\end{figure}

In the previous section, we presented different \acp{nmie} techniques. These methods can be used to estimate the mutual information by jointly sampling the channel input and the channel output, without knowing the underlying channel models. In this section, we use these \acp{nmie} as part of an optimization process for estimating the channel capacity from data samples. 

The block diagram for neural capacity estimation approach is shown in Fig.~\ref{fig:nceBlock}. In this approach, which was also used in \cite{aharoni20_CapMemChan}, the channel input $X$ is generated from a distribution that is learned by a neural network called the \ac{ndt}, parameterized by $\theta$. 
{\nf Specifically, the \ac{ndt} take random samples $S$, generated from a known distribution $P_S$, as input and outputs $X$ (i.e., samples of the channel input). Given a sufficiently large neural network, which can act as a general function approximator, the distribution of $P_S$ can be transformed into an arbitrary channel input distribution $P_X$, through parameters $\theta$. Note that the activation function at the final layer of the \ac{ndt} should be chosen to match the support of $X$.}
Then, given a channel and the channel input $X$, the channel output $Y$ is generated. Samples from the channel input and output are then used to estimate the mutual information using \acp{nmie}, parameterized by $\phi$. Therefore, the capacity is estimated using a double maximization
\begin{align}
    \label{eq:nce_opti}
    \hat{\Ccal} = \max_\theta \max_\phi I_\phi(f_\theta(S); Y),
\end{align}
where $S$ are samples drawn from a known distribution, $f_\theta$ is the \ac{ndt}, and $I_\phi$ is the \ac{nmie}.

{\nf The \ac{nmie} network and \ac{ndt} are trained iteratively to estimate the mutual information for a given $P_X$ and updating $P_X$ to maximize mutual information. Hence, a single training iteration has two phases. In phase one, the \ac{nmie} is trained exclusively to estimate the mutual information. This is achieved by keeping the weights of the \ac{ndt} network constant and only updating the weights of the \ac{nmie} during backpropagation. In the second phase, the weights of the \ac{nmie} is kept constant and the weights of the \ac{ndt} network are updated using backpropagation. The estimate of the capacity is obtained when this iterative optimization converges or certain number of training steps are reached. 

Note that this method is very similar in sprite to the generative adversarial networks \cite{GAN}, where a generator network transforms random samples from a known distribution to random image samples, and a discriminator network classifies the generated images as real or fake. An iterative optimization process is then employed to improve the distribution over which random images are generated from, and the classification accuracy of the discriminator. At convergence, the output of the discriminator is random as it is unable to identify an image as real or fake.}

\begin{algorithm}[t]
	\caption{Neural capacity and optimal input estimation}
	\label{alg:NCE}
	\KwIn{Channel model or its GAN approximation}
	\KwOut{Estimated capacity and optimal input distribution}
	Initialize parameters of the {NMIE} and \ac{ndt} randomly \\
	\While{not converged or max iteration not reached}
	{
		\textbf{Phase 1}: \textbf{Train {NMIE}} \\
		Generate $B$ sample of $S$: $\{s_{i}\}_{i=1}^B$ \\ 
		Generate $\{(x_i, y_i)\}_{i=1}^B$ using \ac{ndt} and the channel \\
		Calculate the MI loss in \eqref{eq:nce_opti} with respect to $\phi$ $$\mathcal{L(\phi)} = - I_\phi(f_\theta(S), Y)$$ 
	    Train the {NMIE} $I_\phi$ using Adam optimizer \\
	    
		\textbf{Phase 2}: \textbf{Train NDT}\\
		Generate $B$ sample of $S$: $\{s_{i}\}_{i=1}^B$ \\ 
		Generate $\{(x_i, y_i)\}_{i=1}^B$ using \ac{ndt} and the channel \\
		Calculate the MI loss in \eqref{eq:nce_opti} with respect to $\theta$ $$\mathcal{L(\theta)} = - I_\phi(f_\theta(S), Y)$$ 
		Train NDT network $f_\theta$ using Adam optimizer
	}
	
	\textbf{Return}: $\hat{\Ccal}=I_\phi(X;Y)$ and $\hat{p}_\theta$	using large samples of $S$ drawn at random.
\end{algorithm}
\begin{algorithm}
\caption{Neural capacity estimation for channels with discrete optimal input.}
\label{alg:NCE_discrete}
\KwIn{Channel model or its GAN approximation}
\KwOut{Estimated capacity and optimal input distribution}
Set the number mass points $m=2$ and $\hat{\Ccal}_1 = 0$  \\
Use Algorithm~\ref{alg:NCE} with $p_S(s)$ having $m$ uniform mass points in $[0,m-1]$ \\
Set $\hat{\Ccal}_{m}$ to be the output of the Algorithm~\ref{alg:NCE}

\While{$\hat{\Ccal}_{m} > \hat{\Ccal}_{m-1}$} 
{ $\hat{\Ccal}_{m-1} = \hat{\Ccal}_{m}$\\
increment the number mass points $m=m+1$\\
Use Algorithm~\ref{alg:NCE} with $p_S(s)$ having $m$ uniform mass points in $[0,m-1]$\\
Set $\hat{\Ccal}_{m}$ to be the output of the Algorithm~\ref{alg:NCE} \\
}
\textbf{Return}: $\hat{\Ccal}_{m-1}$ and the corresponding learned input distribution
\end{algorithm}

Algorithm~\ref{alg:NCE} summarizes this procedure. Several factors affect the performance of the neural capacity estimator, namely the choice of \ac{nmie} technique from Section \ref{sec:NeuralMIestimator}, and the random distribution used for generating the samples $s_i$. When the optimal input distribution is known to be continuous, the normal Gaussian distribution can be a good choice for $P_S$. When the channel has discrete outputs (like Poisson channel), and optimal input is known to have discrete mass points, we use Algorithm~\ref{alg:NCE_discrete}. The reason behind this change in the algorithm for computing capacity is that we have found experimentally that using continuous normal Gaussian distribution for channels with discrete output and discrete optimal distributions have a degraded performance with Algorithm~\ref{alg:NCE} compared to using discrete distributions as initialization distributions.

This {algorithm} starts by evaluating the capacity using Algorithm~\ref{alg:NCE} with a uniform binary distribution for $P_S$. Then a new mass point is added to the distribution of $S$ and the capacity is re-evaluated using Algorithm~\ref{alg:NCE}. This process of adding new mass points to the distribution of $S$ and then re-evaluating capacity is continued until the estimated capacity no longer improves due to the newly added mass point. As we show in our numerical results, this proposed approach can achieve better estimate of the capacity for channels with discrete mass points such as the Poisson channel.   

One of the advantages of using neural capacitor estimators compared to other methods like Blahut-Arimoto or deterministic annealing is their running time. In fact, while neural capacity estimators converge in the order of minutes to the actual capacity in different SNR ranges for different channels, other methods such as Blahut-Arimoto need several hours or days to converge to the capacity value (even in simple channel models with limited alphabets). 

Despite this advantage, neural capacity estimation can be unstable based on the choice of distribution for $S$, and the choice of \ac{nmie} technique. Therefore, in this work, we evaluate the performance of neural capacity estimation using different \ac{nmie} methods, and different distributions for $S$ for variety of communication channels. To the best of our knowledge this is the most comprehensive empirical study of neural capacity estimators. In the next section, we describe the channel models we will consider as part of our empirical study, followed by our numerical results.    

\subsubsection{Implementation details}
Let us briefly introduce here the NDT and NMIE architecture, training, and simulations.
All DNN consists of multi-layer fully-connected neural networks, with up to four layers. 
The batch size for training is $256$, and we use unit variance Gaussian distribution as initialization, unless otherwise stated.
For obtaining a histogram of the input distribution, we use  $64000$ samples. 
The ADAM  optimizer is used in all cases with a learning rate=0.0001.

\section{Channel Models}
\label{sec:Channel Models and Related Results}

In the remainder of the paper, we investigate the performance of the neural capacity estimators and the quality of the learned input distribution. 
For this purpose, we compare the performance of the proposed neural estimator with the capacity results derived in the literature.
%
%
Before we present our numerical results, in this section, we present the channel models studied in the reminder of the paper and present 
relevant bounds on capacity, and characterizations of the corresponding capacity-achieving input distribution.

\subsection{Point-to-Point (P2P) Memoryless Channels}
\label{Point-to-Point  Memoryless Channels}

The point-to-point (P2P) channel describes the communication between one transmitter and one receiver as described by the conditional probability $P_{Y|X}$.
For the P2P class of channels, it has been show  \cite{shannon1948mathematical,verdu1994general} that the capacity  is obtain by the supremum of the mutual information between input and output over all feasible input distributions, i.e.
\ea{
\Ccal(P_X) = \sup_{P_X} I(X,Y).
}
Let us briefly introduce the models we shall investigate in the remainder of the paper.

\subsubsection{Additive White Gaussian Noise (AWGN) Channel}
\label{sec:Additive White Gaussian Noise (AWGN) channel}

The AWGN channel \cite{shannon1948mathematical} is perhaps the most celebrated and widespread communication model.
In this model the input/output relationships is obtained as 
\ea{
Y^N = X^N + Z^N,
\label{eq:additive noise}
}
where $N$ is the block-length, $Z^N$ is a vector of i.i.d standard normal distribution, and  the input is subject to the average power constraint
\ea{
\sum_{n \in [N]} \| X_n\|_2^2  \leq NP.
\label{eq:second moment}
}
The capacity of the AWGN channel is well-known and given by:
\ea{
\Ccal_{\rm AWGN}(P) = \f 12 \log \lb 1+ P\rb.
\label{eq:c awgn}
}
Note that the LHS of \eqref{eq:c awgn} explicitly captures the dependency on the second moment from the constraint in \eqref{eq:second moment}.
Additionally, it is well-known that the capacity achieving input distribution is the Gaussian distribution with zero mean and variance $P$.
%

\subsubsection{Optical Intensity (OI) Channel}
\label{sec:Optical intensity channel}
%
%

For the OI channel, the input/output relationship is again given by \eqref{eq:additive noise}, but the input is subject to the constraints
\begin{align}
  \Pr[ X_n > A \vee X_n < 0]  & =0, \,\,\,\,
\mathbb{E}[X_n] \leq  \Ecal,
\label{eq:pos constraint}  
\end{align}
\noindent for all $n \in [N]$. As in  \cite{5238736}, we denote the ratio between the average power and
the peak power as  
\ea{
\al = \f {\Ecal}{A}.
}
%
Note that, while the input is constrained to be positive, there is no restriction on output of channel, which could be either positive or negative.

The capacity of the OI channel is indicated as $\Ccal_{OI}(A,\Ecal)$, where again we have explicitly indicated the dependency of the constraints in \eqref{eq:pos constraint}. 
An analytical expression for the capacity of the OI channel $\Ccal_{OI}(A,\Ecal)$ is not known, however, prior works have found upper and lower bounds on capacity \cite{hra04_boundOI, far10_boundOI, cha16_boundOI,5238736}.  We present the best known bounds on the capacity of the OI channel, used in this work for the performance evaluation  in the theorem below for convenience.

 \begin{theorem}[Bounds on the capacity of the OI channel \cite{5238736}]
\label{thm:OI}
The capacity of the OI channel for $1/2 \leq\alpha\leq 1$ is lower-bounded as
 \begin{equation}
  \Ccal_{\rm OI}(A,\alpha A)\geq  \f 12  \log \lb 1+\frac{A^{2}}{2\pi e \sigma^{2}} \rb.
 \end{equation}
 and upper-bounded as one equations in  \cite[eq.19]{5238736} and \cite[eq.20]{5238736} (the smallest determines the upper bound).
%
\end{theorem}

\subsubsection{Peak Power-Constrained AWGN  (PPC-AWGN) Channel}
\label{sec:PPC-AWGN Channel}

The PPC-AWGN is again obtained from the Gaussian additive P2P channel in \eqref{eq:additive noise} but under the additional peak-power constraint 
\ea{
\| X_n\|_{\infty} \stackrel{\tiny \rm a.s. }{\leq}  A \quad \forall\ n \in [N].
}

%

A closed-form expression for the capacity for the PPC-AWGN channel, $\Ccal(P,A)$, and the corresponding optimal input distribution are not know in general. Therefore, in this work, we use the lower and upper bounds on the capacity derived in \cite{kellips} for the performance evaluation of the neural capacity estimators. The specific bounds used for evaluation are presented in the theorems below for convenience.

\begin{theorem}[Bounds on the capacity of PPC-AWGN \cite{kellips}]
\label{thm:PPC}
The  capacity of the PPC-AWGN channel for $A=\sqrt{P}$ is upper-bounded as 
\ea{
\Ccal_{\rm PPC}(P)  \leq \min \lcb \log\lb  1+\sqrt{\frac{2P}{\pi e}}\rb,\f 12 \log(1+P) \rcb,
}
and lower-bounded as
\ea{
\Ccal_{\rm PPC}(P) \geq -\int_{-\infty}^{\infty}f(y)\log(f(y))dy -\f 12 \log(2\pi e),
}
where $f(y)$ is the mixture density resulting from the sum of the
discrete $M$-point uniform input distribution with independent
unit-variance Gaussian noise.
\end{theorem}

A refinement of the upper bound from \cite{kellips} is derived in \cite{kramer} for some specific intervals of $P$. 

\begin{theorem}[Upper-bound on the capacity of PPC-AWGN \cite{kramer}]
\label{thm:PPC 2}
The  capacity of the PPC-AWGN channel 
for $A=\sqrt{P}$ is 
upper-bounded as
\ea{
\Ccal_{\rm PPC}(P) \leq \beta(P)\log \lb \frac{2P}{\pi e}\rb+H_{2}(\beta(P)),
}
for $P_{\rm dB} \in [2:6] \  \rm{dB}$  and $\beta(P)=\f 12-\Qcal(2\sqrt{P})$.

\end{theorem}

\subsubsection{Poisson Channel}
\label{sec:Poisson Channel}

The Poisson channel has a formulation that is similar to the OI channel in which the channel input $X_n$ is continuous over the interval $[0, A]$. Unlike the OI channel, the channel output $Y_n$, however, is discrete and positive. 
More specifically, the Poisson channel is described by the conditional probability
\ea{
&p_{Y_n|X_n}(y_n \mid x_n)=\frac{e^{-(x_n+\lambda_0)}(x_n+\lambda_0)^{y_n}}{y_n!}, 
\label{eq:poisson in/out}
}
for $y_n \in \Nbb$ and where $\lambda_0$ is noise due to dark current. The channel input has a peak constraint $A$, and an average power constraint
\ea{
\sum_{n \in [N]}  X_n  \leq \Ecal N.
}
We denote the capacity of this channel as $\Ccal_{\rm PC}(\Ecal, A)$.
An analytical expression for $\Ccal_{\rm PC}(\Ecal, A)$  is not known in general.
In \cite{shamai1}, it was shown that the capacity-achieving input distribution has finite number of mass points and the number of mass points goes to infinity as the constraints are relaxed.
Upper and lower bounds on the capacity for different input constraints were derived in \cite{moser}. Finally, numerical methods such as deterministic annealing can be used to numerically evaluate capacity and the optimal input \cite{caocapacity11}.

In the numerical results, we use the lower bounds and upper bounds from \cite{moser} to evaluate the performance of neural capacity estimators on the Poisson channel. The bounds are restated as theorems below for convenience.

\begin{theorem}[Th. 3-\cite{moser}]
\label{th:poisson}
When 
\ea{
 0 \leq  \frac{\Ecal}{A}\leq \frac{1}{3},
 }
the capacity of the Poisson channel is  lower-bounded as \cite[eq.12]{moser}.
Additionally, the capacity can be upper-bounded as \cite[eq.13-eq.16]{moser}.

\end{theorem}
Another theorem in \cite{moser} bounds the capacity in the ${1}/{3}\leq {\Ecal}/{A}\leq 1$ regime.

\begin{theorem}{(Th. 4-\cite{moser})}
When $\frac{1}{3}\leq\frac{\Ecal}{A}\leq 1,$
the capacity of the Poisson channel is lower-bounded as \cite[eq.18]{moser}
Additionally, the capacity in this range for ${\Ecal}/{A}$ can be upper-bounded as \cite[eq.19]{moser}.
\end{theorem}

It should be noted that upper bound and lower bound above are asymptotically tight but are not tight over the whole SNR range.
The authors of \cite{chen1} propose the Deterministic Annealing (DA) algorithm for the evaluation of the capacity of the Poisson channel. We will also compare the neural capacity estimators with this approach.

\subsection{Multiple Access Channel (MAC) }
\label{sec:Multiple Access (MAC) Channel}

The 2-user MAC channel is the multi-terminal channel model in which two users simultaneously and synchronously transmit toward a common receiver. 
The relationship between the input at the two transmitters and the output at the receiver is described by the transition probability $P_{Y|X_1,X_2}$.
For this model, the ultimate communication performance under which reliable communication is possible is described by the \emph{capacity region}, that is a tuple of transmission rates $(R_1,R_2)$ for which error probability can be made vanishing small for large block lengths.

In \cite{liao1972coding,ahlswede1973multi} it was shown that the capacity of the MAC is obtained as the convex hull of rate points satisfying
\eas{
R_1  & \leq I(Y;X_1|X_2)\\ 
R_2  &  \leq I(Y;X_2|X_1), \\
R_1+R_2  \leq I(Y;X_1,X_2),
}{\label{eq: MAC C}}
for some $P_{X_1,X_2}=P_{X_1}P_{X_2}$.

\subsubsection{AWGN MAC}
\label{sec:AWGN MAC channel}

Among the most studied MAC models, AWGN MAC is perhaps the most well-known model. 
In this channel, the input/output relationship can be expressed as 
\ea{
Y^N=X_1^N + X_2^N+Z^N,  
\label{eq:awgn mac}
}
where $X_1^N$ and $X_2^N$ are the channel inputs from users 1 and 2, respectively, and $Z^N$ is an i.i.d. Gaussian sequence with zero mean and unitary variance.
The input is further subject to a power constraint analogous to that in \eqref{eq:second moment} as 
\ea{
\sum_{n \in [N]} \| X_{i,n}\|_2^2  \leq NP_i.
\label{eq:second moment MAC}
}
for $i \in \{1,2\}$.

%
The convex hull of the tuple in \eqref{eq: MAC C}
 can be shown to be equivalent to the following region
\eas{
R_1 & \leq \f 12 \log (1+P_1)\\ 
R_2 & \leq \f 12 \log (1+P_2)\\
R_1+R_2 & \leq \f 12 \log (1+P_1+P_2).
}{\label{eq: awgn MAC C}}
For the model in \eqref{eq:awgn mac}, it is further known that the capacity achieving input distribution is Gaussian, that is, any tuple on the boundary of  \eqref{eq: awgn MAC C} can  be achieved through Gaussian distributed inputs at both transmitters. 

\subsubsection{Optical Intensity (OI) MAC}
\label{sec:Optical Intensity (OI) MAC channel}

Similarly to the relationship between the AWGN and the OI channel,  in Sec.  \ref{sec:Additive White Gaussian Noise (AWGN) channel}
and Sec. \ref{sec:Optical intensity channel} respectively, the OI MAC 
\cite{kahn1997wireless} is described by the
input/output relationship in \eqref{eq:awgn mac} but the channel inputs are subject to the constraint
\eas{
& 0 \leq X_n \leq A_i\\
& \sum_{n \in [N]}X_i \leq N \Ecal_i.
\label{eq:oi bounds mean}
}{\label{eq:oi bounds}}
When only the constraint in \eqref{eq:oi bounds mean} is considered,  the capacity region becomes \cite{oimac1},\cite{oimac2}
\eas{
R_1 & \leq \f 12 \log \lb \frac{e}{2\pi}(\Ecal_1+2)^{2} \rb\\
R_2  & \leq \f 12 \log \lb\frac{e}{2\pi}(\Ecal_2+2)^{2}\rb \\
R_1+R_2  & \leq \f12 \log \lb \frac{e}{2\pi} (\Ecal_1+\Ecal_2+2)^{2} \rb.
}{\label{eq:oi mac c}}
The channels described in this section results will be used in the next sections to evaluate the performance of the neural capacity estimators as well as the quality of the learned input distribution.

 \section{AWGN channel: Bench-marking NMIEs}
\label{sec:Comparison existing results}

In this section, we benchmark the MI estimators discussed in Sec. \ref{sec:NeuralMIestimator} by comparing the estimated value of capacity and the learned input distribution to the optimal ones for the AWGN channel described in Sec. \ref{sec:Additive White Gaussian Noise (AWGN) channel}.
Since a closed-form expression for both the capacity and optimal input distribution is available for the AWGN channel model, this simple example provides a number of insights on the choice of \ac{nmie}.

Results for this channel are obtained using 10 separate estimation trials, where for each trial the \ac{ndt} and the \ac{nmie} networks are initialized randomly. 
In Table \ref{tab:AWGN}, we provide a comparison between the true capacity and its neural estimate for the $2\rm dB$, $20 \rm dB$, and $40 \rm dB$ SNRs. 
Specifically, we provide the average estimate across the 10 trials, and the same number of trials for variance of the capacity estimation.
We note that the relative accuracy decreases as the SNR grows but the loss of accuracy does not affect all methods equally.
The MINE method gives a more accurate estimate of the capacity compared to other methods, while \ac{dine} seems to diverge from the true capacity at higher SNRs.  Moreover, the standard deviation of MINE is lower than all other methods suggesting that it is more robust to random initialization and has a lower variance in estimating capacity. Note that this is contrary to estimation of mutual information, where methods such as \ac{smile} show lower variance.      

Fig.~\ref{fig:hist} shows the histogram of the \ac{ndt} 
learned input distribution for the \ac{dine}, \ac{mine}, and \ac{smile}. We observe that at higher SNRs the \ac{mine} achieves a more accurate approximation of the optimal input distribution, i.e. Gaussian. While \ac{dine} can also learn accurate input distributions at low SNRs, at higher SNRs, it seems to deviate from the optimal input. Note that all three methods are trained for the same number of iterations here for a fair comparison. 

During testing, we note that the clipping operation used in \ac{smile},  significantly effects the quality of the learned input distribution. After experimenting with the range of values in the range $\tau \in (10,0)$ for $\clip$ in \eqref{eq:smile bound}, we have found clipping value $\tau=0.2$ gives the best estimates of capacity in terms of stability of results.
%
%
%

{\fm \begin{rem}
During testing, we note that the clipping operation used in \ac{smile},  significantly effects the quality of the learned input distribution. After experimenting with the range of values in the range $\tau \in (10,0)$ for $\clip$ in \eqref{eq:smile bound}, we have found clipping value $\tau=0.2$ gives the best estimates of capacity in terms of stability of results.
Thus, we use this parameters in our all experiments.
\end{rem}}

\begin{table}[b]
\small
\centering
\tabcolsep=0.11cm
{\fm \begin{tabular}{ |c|c|c|c|c|c|c|c|} 
 \hline
  SNR ($\rm dB$) & \ac{dine}  ($\nats$)& \ac{mine}  ($\nats$) & \ac{smile} ($\nats$) & DV variant ($\nats$)& $\chi^{2}$ ($\nats$)&NWJ ($\nats$)& True \\ 
  \hline
2 & $0.39 \pm 0.12$ & $\mathbf{0.48\pm 0.05}$ & $0.42\pm 0.09$&$0.49\pm0.011$& $0.39\pm.051$ &$0.47\pm 0.0017$& 0.474  \\ 
 \hline
20  & $2.21 \pm 0.16$ & $\mathbf{2.30\pm 0.005}$ & $2.12\pm 0.07$&$2.29\pm0.021$&$2.19\pm.062$ &$2.30\pm0.008$& 2.307  \\ 
 \hline
 40 & $4.11\pm 0.23$  &  $\mathbf{4.51\pm 0.08}$ & $4.42\pm 0.17$&$4.41\pm.15$& $4.101\pm0.12$&$4.49\pm 0.037$  & 4.605  \\ 
 \hline
\end{tabular}}
\caption{Estimated capacity of AWGN channel using different methods as estimated across  10 trials.}
\label{tab:AWGN}
\end{table}

 \begin{figure*}[t!]
   \centering
  \begin{tikzpicture}
    \definecolor{mycolor1}{rgb}{0.00000,0.44706,0.74118}%
    \definecolor{mycolor2}{rgb}{0.63529,0.07843,0.18431}%
    \definecolor{mycolor3}{rgb}{0.00000,0.49804,0.00000}%
    
    \begin{groupplot}[
        group style={
            group name=my plots,
            group size=3 by 6,
            xlabels at=edge bottom,
            xticklabels at=edge bottom,
            vertical sep=0pt,
            horizontal sep=5pt
        },
        height=4.5cm,
        width=6.5cm,
        ymin=0,
        ytick=\empty,
        xmajorgrids,
        ymajorgrids,
     ]
]


\nextgroupplot[ylabel=DINE,xtick={-4,-3,...,4}]
\coordinate (top) at (axis cs:1,\pgfkeysvalueof{/pgfplots/ymax});

\addplot[red,samples=100,domain=-4:4,samples=500,fill=red!120, opacity=0.4]{(1/(1.25*sqrt(2*pi)))*exp(-(x^2/(2*1.58)))};

\addplot+[ybar interval,mark=no,fill=blue!120,draw=blue, opacity=0.5, ]
    table[col sep=comma]{./data_m/data_batch_hs-dine-awgn-2.txt};

\nextgroupplot
[xtick={-30,-20,...,30},]

\coordinate (top) at (axis cs:1,\pgfkeysvalueof{/pgfplots/ymax});
 
 \addplot[red,domain=-30:30,samples=500,fill=red!120, opacity=0.4]{(1/(10*sqrt(2*pi)))*exp(-(x^2/(2*100)))};
 
 \addplot [ybar interval,mark=no,fill=blue!120,draw=blue, opacity=0.5]
 [restrict x to domain=-30:30]
 table[col sep=comma]{./data_m/data_batch_hs-dine-awgn-20.txt };
 
\nextgroupplot[xtick={-500,-250,...,500}]

\coordinate (top) at (axis cs:1,\pgfkeysvalueof{/pgfplots/ymax});

\addplot[red,domain=-500:500,samples=500,fill=red!120, opacity=0.4]
%
{(1/(100*sqrt(2*pi)))*exp(-(x^2/(2*10000)))};
\addplot[ybar interval,mark=no,fill=blue!120,draw=blue, opacity=0.5,] 
table[col sep=comma]{./data_m/data_batch_hs-dine-awgn-40.txt};

\nextgroupplot[ylabel=MINE,xtick={-4,-3,...,4}]

    \addplot[red,samples=100,domain=-4:4,samples=500,fill=red!120, opacity=0.4]{(1/(1.25*sqrt(2*pi)))*exp(-(x^2/(2*1.58)))};
\addplot+[ybar interval,mark=no,fill=blue!120,draw=blue, opacity=0.5]
  [restrict x to domain=-4:4]
    table[col sep=comma]{./data_m/data_batch_hs-mine-awgn-2.txt};

\nextgroupplot[xtick={-30,-20,...,30}]

\coordinate (top) at (axis cs:1,\pgfkeysvalueof{/pgfplots/ymax});
 
 \addplot[red,domain=-30:30,samples=500,fill=red!120, opacity=0.4]{(1/(10*sqrt(2*pi)))*exp(-(x^2/(2*100)))};

 \addplot [ybar interval,mark=no,fill=blue!120,draw=blue, opacity=0.5]
 [restrict x to domain=-30:30]
 table[col sep=comma]{./data_m/data_batch_hs-mine-awgn-20.txt};
 
\nextgroupplot[xtick={-500,-250,...,500}]

\addplot[red,domain=-500:500,samples=500,fill=red!120, opacity=0.4]
%
{(1/(100*sqrt(2*pi)))*exp(-(x^2/(2*10000)))};
\addplot[ybar interval,mark=no,fill=blue!120,draw=blue, opacity=0.5,] 
table[col sep=comma]{./data_m/data_batch_hs-mine-awgn-40.txt};


\nextgroupplot[ylabel=SMILE,xtick={-4,-3,...,4},]

\addplot[red,samples=100,domain=-4:4,samples=500,fill=red!120, opacity=0.4]{(1/(1.25*sqrt(2*pi)))*exp(-(x^2/(2*1.58)))};

\addplot+[ybar interval, mark=no,
            fill=blue!120,draw=blue, opacity=0.5,]
            [restrict x to domain=-4:4]
    table[col sep=comma]{./data_m/data_batch_hs-smile-awgn-2.txt};
\coordinate (bot) at (axis cs:1,\pgfkeysvalueof{/pgfplots/ymin});

\nextgroupplot[xtick={-30,-20,...,30} ]

\coordinate (top) at (axis cs:1,\pgfkeysvalueof{/pgfplots/ymax});

 \addplot[red,domain=-30:30,samples=500,fill=red!120, opacity=0.4]{(1/(10*sqrt(2*pi)))*exp(-(x^2/(2*100)))};

 \addplot [ybar interval,mark=no,fill=blue!120,draw=blue, opacity=0.5]
  [restrict x to domain=-30:30]
 table[col sep=comma]{./data_m/data_batch_hs-smile-awgn-20.txt};

\coordinate (bot) at (axis cs:1,\pgfkeysvalueof{/pgfplots/ymin});

\nextgroupplot[xtick={-500,-250,...,500}]

\addplot[red,domain=-500:500,samples=500,fill=red!120, opacity=0.4]
%
{(1/(100*sqrt(2*pi)))*exp(-(x^2/(2*10000)))};
\addplot[ybar interval,mark=no,fill=blue!120,draw=blue, opacity=0.5,] 
table[col sep=comma]{./data_m/data_batch_hs-smile-awgn-40.txt};

\nextgroupplot[ylabel=DV varainat]

\addplot[red,samples=100,domain=-4:4,samples=500,fill=red!120, opacity=0.4]{(1/(1.25*sqrt(2*pi)))*exp(-(x^2/(2*1.58)))};

\addplot+[ybar interval, mark=no,
            fill=blue!120,draw=blue, opacity=0.5,]
            [restrict x to domain=-4:4]
    table[col sep=comma]{./data_m/data_batch_dv_var_2db.txt};
\coordinate (bot) at (axis cs:1,\pgfkeysvalueof{/pgfplots/ymin});

\nextgroupplot[xtick={-30,-20,...,30},]

\coordinate (top) at (axis cs:1,\pgfkeysvalueof{/pgfplots/ymax});

 \addplot[red,domain=-30:30,samples=500,fill=red!120, opacity=0.4]{(1/(10*sqrt(2*pi)))*exp(-(x^2/(2*100)))};

 \addplot [ybar interval,mark=no,fill=blue!120,draw=blue, opacity=0.5]
  [restrict x to domain=-30:30]
 table[col sep=comma]{./data_m/data_batch_dv_var_20db.txt};

\coordinate (bot) at (axis cs:1,\pgfkeysvalueof{/pgfplots/ymin});

\nextgroupplot[xtick={-500,-250,...,500}]

\addplot[red,domain=-500:500,samples=500,fill=red!120, opacity=0.4]
%
{(1/(100*sqrt(2*pi)))*exp(-(x^2/(2*10000)))};
\addplot[ybar interval,mark=no,fill=blue!120,draw=blue, opacity=0.5,] 
table[col sep=comma]{./data_m/data_batch_dv_var_40db.txt};
\nextgroupplot[ylabel=Chi-bound,xtick={-4,-3,...,4},]

\addplot[red,samples=100,domain=-4:4,samples=500,fill=red!120, opacity=0.4]{(1/(1.25*sqrt(2*pi)))*exp(-(x^2/(2*1.58)))};;

\addplot+[ybar interval, mark=no,
            fill=blue!120,draw=blue, opacity=0.5,]
            [restrict x to domain=-4:4]
    table[col sep=comma]{./data_m/data_batch_chi_2.txt};;

\nextgroupplot[xtick={-30,-20,...,30}]

\coordinate (top) at (axis cs:1,\pgfkeysvalueof{/pgfplots/ymax});

 \addplot[red,domain=-30:30,samples=500,fill=red!120, opacity=0.4]{(1/(10*sqrt(2*pi)))*exp(-(x^2/(2*100)))};

 \addplot [ybar interval,mark=no,fill=blue!120,draw=blue, opacity=0.5]
  [restrict x to domain=-30:30]
 table[col sep=comma]{./data_m/data_batch_chi_20.txt};

\coordinate (bot) at (axis cs:1,\pgfkeysvalueof{/pgfplots/ymin});

\nextgroupplot[xtick={-500,-250,...,500}]

\addplot[red,domain=-500:500,samples=500,fill=red!120, opacity=0.4]
%
{(1/(100*sqrt(2*pi)))*exp(-(x^2/(2*10000)))};
\addplot[ybar interval,mark=no,fill=blue!120,draw=blue, opacity=0.5,] 
table[col sep=comma]{./data_m/data_batch_chi-40.txt};
\nextgroupplot[ylabel=NWJ, xlabel=  $2 \rm dB$,xtick={-4,-3,...,4},]

\addplot[red,samples=100,domain=-4:4,samples=500,fill=red!120, opacity=0.4]{(1/(1.25*sqrt(2*pi)))*exp(-(x^2/(2*1.58)))};;

\addplot+[ybar interval, mark=no,
            fill=blue!120,draw=blue, opacity=0.5,]
            [restrict x to domain=-4:4]
    table[col sep=comma]{./data_m/data_batch_hs_nwj_2.txt};

\nextgroupplot[xtick={-30,-20,...,30}, xlabel=$20 {\rm dB}$ ]

\coordinate (top) at (axis cs:1,\pgfkeysvalueof{/pgfplots/ymax});

 \addplot[red,domain=-30:30,samples=500,fill=red!120, opacity=0.4]{(1/(10*sqrt(2*pi)))*exp(-(x^2/(2*100)))};

 \addplot [ybar interval,mark=no,fill=blue!120,draw=blue, opacity=0.5]
  [restrict x to domain=-30:30]
 table[col sep=comma]{./data_m/data_batch_nwj_20.txt};

\coordinate (bot) at (axis cs:1,\pgfkeysvalueof{/pgfplots/ymin});

\nextgroupplot[xlabel=$40{\rm dB}$,xtick={-500,-250,...,500}]

\addplot[red,domain=-500:500,samples=500,fill=red!120, opacity=0.4]
%
{(1/(100*sqrt(2*pi)))*exp(-(x^2/(2*10000)))};
\addplot[ybar interval,mark=no,fill=blue!120,draw=blue, opacity=0.5,] 
table[col sep=comma]{./data_m/data_batch_hs_nwj_40.txt};
\end{groupplot}
\end{tikzpicture}
   \caption{Comparison of different methods for computing the optimal input distribution for Gaussian channel. The red distribution is the optimal input distribution and the blue histograms are the learned input distribution.}
  \label{fig:hist}
 \end{figure*}
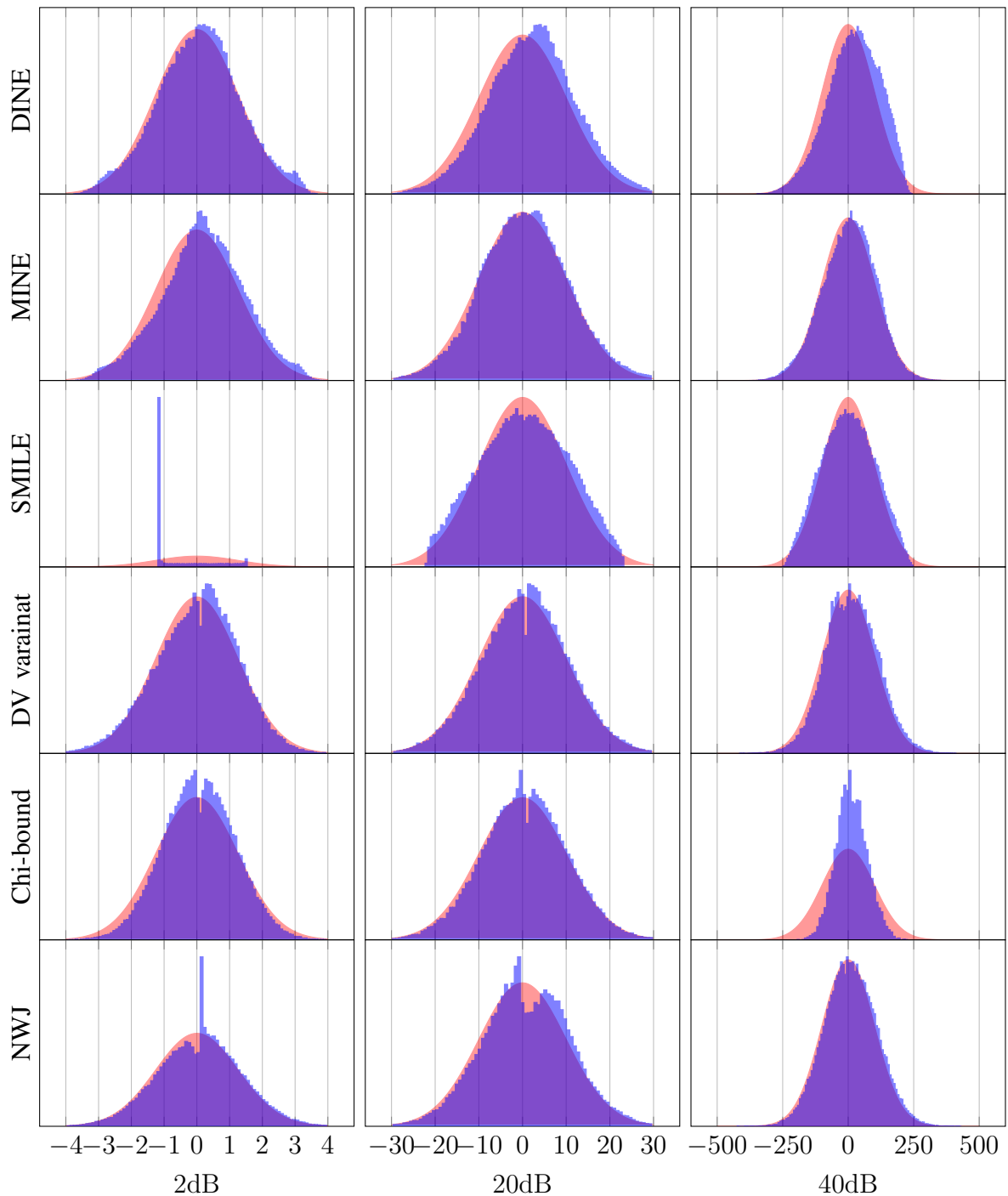
\begin{figure}
    \centering
    \begin{tikzpicture}
  \begin{groupplot}[group style={group size=4 by 1,xlabels at=edge bottom,
            xticklabels at=edge bottom,
            vertical sep=0pt,
            horizontal sep=5pt,},yticklabels=\empty,height=5cm,width=5cm]

    \nextgroupplot
        [title=Sample Size-512]

\addplot+[ybar interval, mark=no,
            fill=blue!120,draw=blue, opacity=0.5,]
            [restrict x to domain=-10:10]
    table[col sep=comma]{./data_m/data_batch_test-hist-2.txt};
    
    \nextgroupplot
    [title=Sample Size=5120]
%

\addplot+[ybar interval, mark=no,
            fill=blue!120,draw=blue, opacity=0.5,]
            [restrict x to domain=-10:10]
    table[col sep=comma]{./data_m/data_batch_test-mine-20.txt};
    \nextgroupplot
    [title=Sample Size-51200]
%

\addplot+[ybar interval, mark=no,
            fill=blue!120,draw=blue, opacity=0.5,]
            [restrict x to domain=-10:10]
    table[col sep=comma]{./data_m/data_batch-test-mine-200.txt};
    \nextgroupplot
    [title=Sample Size-512000]
%

\addplot+[ybar interval, mark=no,
            fill=blue!120,draw=blue, opacity=0.5,]
            [restrict x to domain=-10:10]
    table[col sep=comma]{./data_m/data_batch-test-mine-2000.txt};    
  \end{groupplot}
\end{tikzpicture}
    \caption{Effect of test sample size on the learned input distribution by \ac{ndt} networks. This experiment is done with MINE method, and AWGN channel with average power constraint at 8 $dB$.}
    \label{fig:my_label1}
\end{figure}
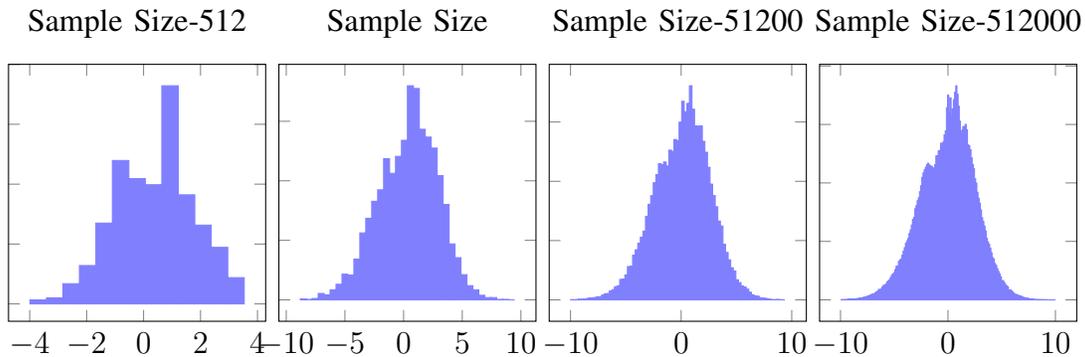


From our extensive simulations, we further glean the following insights, 

\begin{itemize}
    \item 
In moderate values of SNR between, e.g. $P_{\rm dB} \in [5,20]$, all of the methods in Sec. \ref{sec:directMIestimator},
including the SMILE, MINE, DINE and $\chi^{2}$, 
yield tight bounds on capacity.
\begin{table*}
\small
\centering
\begin{tabular}{ |c|c|c|c|c| } 
 \hline
  Test seed-size &$512$& $5120$ & $51200$ & $512000$\\ 
 \hline
 Estimated capacity-MINE & $0.966\pm 0.0012 $& $0.981\pm 0.0021 $& $0.988\pm 0.0019$ &$0.994\pm 0.0024$\\
 \hline
 Estimated capacity-SMILE & $1.06\pm 0.002$& $1.04\pm 0.002$& $1.02\pm 0.002$&$1.001\pm 0.002 $\\
 \hline
 Estimated capacity-DINE & $0.9512\pm 0.005$&$0.97\pm 0.0061$&$0.97\pm 0.004$&$0.981\pm 0.006$\\
 \hline
 Estimated capacity-DV-variant &$0.965\pm .002$&$0.971\pm .003$&$0.985\pm 0.003$&$.989\pm 0.004$\\
 \hline
 Estimated capacity-Chi-Square & $0.945\pm 0.001$&$0.954\pm 0.002$&$0.969\pm 0.002$&$0.981\pm 0.002$\\
 \hline
 Estimated capacity-NWJ& $0.9701\pm 0.001$ &$0.977\pm 0.002$& $0.982\pm 0.003$&$.992\pm 0.005$\\
 \hline
 True capacity & $0.9945$ &$ 0.9945$ &$0.9945$ &$0.9945$\\
 \hline
\end{tabular}
\caption{Effect of number of test samples size on the capacity estimation for different methods over AWGN channel at $8 \db$. The results are in $\nats$.  
}
\end{table*}
\item 
%
In high SNR regime, e.g. $P_{\rm dB} \in [20,\infty ]$,  DINE method fails to capture the Gaussian shape of optimal input while giving poor estimates compared to SMILE and MINE. 
In this regime, NWJ is a close second to MINE. 

\item
Finally, although \ac{smile}
method gives results with lower variance than the other methods, 
MINE  and the NWJ method reach more accurate estimates compared to  other methods.
%

\end{itemize}

\begin{rem}
We have not included the classifier-based for mutual information method and the InfoNCE bound in the numerical results. The reason is after extensive experiments we find that in the joint optimization setting these methods has a very large variance and in many cases it does not converge to the actual capacity values for channel even in the simplest Gaussian case.
For instance the classifier-based approach fails in providing an accurate estimate event for the AWGN channel even in the rather simple case of $SNR = 10 \db$. In this case the estimated capacity is $0.18$, while the true capacity is $1.19$.
\end{rem}

   \section{NMIE for Capacity Investigation}
\label{sec:Numerical Study}


In Sec. \ref{sec:Comparison existing results} we validated the neural capacity estimation described in Sec. \ref{sec:NeuralMIestimator} by considering a well-understood channel model. 
In this section, we benchmark the \ac{nmie}  for channel models in which no tight capacity bounds exists and for which the optimal input distribution is not well understood.

 From a high level perspective, an overview of the results in this section is as follows. 
%
Although all methods in Sec. \ref{sec:NeuralMIestimator} can be used for obtaining capacity bounds, the \ac{mine} method has the most reliable performance. 
For this reason, most of the results in the section will rely on this method and focus on different aspects of this approach, such as initialization, stability, variance of the estimates. 
To this end, in the first two subsection, we study the  OI channel and PPC-AWGN channel and investigate whether and how accurately \ac{nmie} is capable of identifying the discrete masses in the optimal input distribution.

In Sec. \ref{sec:The Poisson Channel},  we study the Poisson channel and propose a discrete initialization procedure that promotes sparsity in the optimal input distribution estimates.

\subsection{The Optical Intensity (OI) Channel}
\label{sec:Optical intensity channel sims}

%
In this section, let us further explore the merits of the various approaches in Sec. \ref{sec:directMIestimator}   by considering the OI channel in Sec. \ref{sec:Optical intensity channel}.
The capacity of optical intensity is generally unknown \cite{oimac1}, \cite{oimac2}. Since the exact capacity with closed-form solution is not available for this channel we compare our results with the bounds in \cite{oimac1}, which are some of the best known bounds in the literature which we have presented in Th. \ref{thm:OI}.
Table.~\ref{tab:OI} shows the estimated value of the capacity and achievable rates using different \ac{nmie} methods for the optical intensity channel with only the average power constraint. 
All methods converge to values that are between lower and upper bounds. Moreover, on average the estimates of \ac{mine} have lower standard deviation compared to other methods.

%
For this example, we consider a continuous Gaussian distribution for DNN initialization and train the network as in Alg. \ref{alg:NCE}.
The results of the simulation for the various estimators are presented in Table. \ref{tab:OI}.
%
The upper and lower bounds in Table. \ref{tab:OI} are evaluated as in Th. \ref{thm:OI}.
%
%

\textcolor{red}{\begin{table*}
\centering
\caption{Comparisons of the different simulation results and bounds for OI channel with only average power constraint, as discussed in Sec. \ref{sec:Optical intensity channel sims},\cite{5238736}.
}
\label{tab:OI} 
{
\small
\begin{tabular}{ |c|c|c|c|c|c|c| } 
 \hline
  SNR ($\rm dB$) & $3$ &$5$ &$8$ & $10$ & $15$ & $20$ \\ 
 \hline
 \ac{mine} ($\nats$)  &$0.395\pm 0.031$&$0.510\pm 0.072$ &$0.710 \pm .077$& $0.840 \pm 0.085 $ & $1.360 \pm 0.140 $& $1.810\pm 0.095$\\
 \hline
 \ac{smile} ($\nats$)&$0.440\pm 0.019 $&$0.460 \pm 0.096 $ &$ 0.670\pm .091$& $0.850\pm 0.181$ & $1.373\pm 0.109$ & $1.831\pm 0.144$\\ 
 \hline
 \ac{dine} ($\nats$) &$0.378\pm 0.071$& $0.429\pm 0.097$ &$0.690\pm .011$&  $0.880 \pm 0.120$ &$1.310\pm 0.121$ &$1.730 \pm 0.110$\\
 \hline
 NWJ ($\nats$)&$0.42\pm 0.001$& $0.61\pm 0.003$&$0.76\pm 0.006$&$0.92\pm 0.01$&$1.39\pm .02$&$1.91\pm 0.012$\\
 \hline
 Chi-Square bound ($\nats$)&$0.422\pm0.001$&$0.452\pm0.004$&$0.601\pm0.10$&$0.85\pm0.015$&$1.35\pm 0.019$&$1.8\pm 0.03$\\
 \hline
 DV-Variant ($\nats$)&$0.45\pm .001$&$0.59\pm .002$&$0.68\pm 0.03$&$0.97\pm 0.05$&$1.36\pm 0.06$&$1.91\pm .08$\\
 \hline
Lower bound ($\nats$) & $0.270$ & $0.420$ & $0.593$ & $0.830$ & $1.340$ & $1.780$\\
\hline
Upper bound ($\nats$) & $0.830$ & $0.990$ & $1.010$ & $1.480$ & $1.770$ & $2.220$ \\
\hline
\end{tabular}
}
\end{table*}}



%
A significant observation for OI channel is that, unlike other methods, \ac{mine} provide accurate estimates of optimal input distributions for OI channel. It is well known that optimal input distribution for OI channel is discrete\cite{oimac2},\cite{5238736} which agrees by our finding using \ac{ndt} and NMIE.

%
%

As we discussed before since $\chi^{2}$, \ac{smile} and \ac{dine} method only provides estimates of achievable rates, for the approximation of optimal input distribution for the channel, we focus on \ac{mine} method.

\begin{figure}
    \centering

\begin{tikzpicture}
    \definecolor{mycolor1}{rgb}{0.00000,0.44706,0.74118}%
    \definecolor{mycolor2}{rgb}{0.63529,0.07843,0.18431}%
    \definecolor{mycolor3}{rgb}{0.00000,0.49804,0.00000}%
    
    \begin{groupplot}[
        group style={
            group name=my plots,
            group size=2 by 1,
            xlabels at=edge bottom,
            xticklabels at=edge bottom,
            vertical sep=15pt,
            horizontal sep=15pt
        },
        height=4 cm,
        width=9cm,
        yticklabels={,,}
        xmajorgrids=true,
     ]
]


\nextgroupplot[title = {case $\epsilon=A=3 {\rm dB}$},xmin=0,xmax=2.5,xtick={0,0.25,...,2.5},xmajorgrids=true,ymajorgrids=true,]

 \addplot [ybar interval, fill=mycolor1!120,draw=mycolor1]
 table[col sep=comma]{./data_m/data-optical-3db.txt};
 
\nextgroupplot[title = {case $\epsilon=A=8 {\rm dB}$},xmin=0,xmax=6.5,xtick={0,0.5,...,6.5},xmajorgrids=true,ymajorgrids=true,]
%

\addplot [ybar interval,fill=mycolor1!120,draw=mycolor1]
    table[col sep=comma]{./Data/data_optical_8db_peak.txt};

 


\end{groupplot}

\end{tikzpicture}
 \caption{Numerical estimate of the optimal input distribution for the OI channel with $\epsilon=A=3 {\rm dB}$  and 
 $\epsilon=A=8 {\rm dB}$ as discussed in Sec. \ref{sec:Optical intensity channel sims}.}  
    \label{optical-intensity-hist}
    \end{figure}
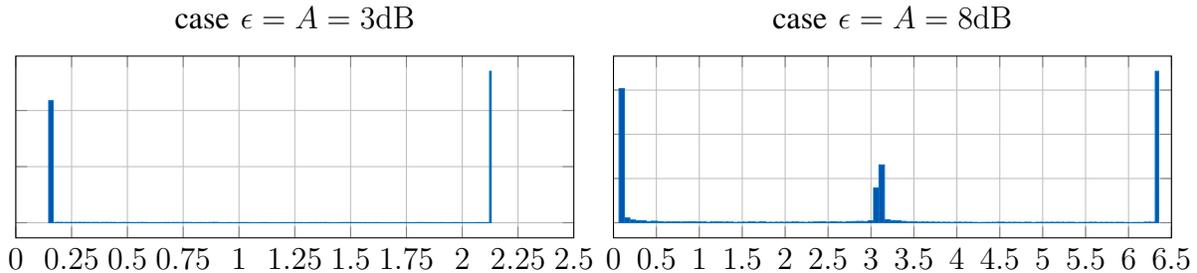
    
In Fig. \ref{optical-intensity-hist} we present the estimated optimal input distribution through \ac{mine} for the case $\epsilon=A$, that is only peak constraint.
For this simulations, the algorithm was initialized using standard Gaussian random variable as initialization input distribution.


\subsection{The Peak Power-Constrained AWGN (PPC-AWGN) Channel}
\label{sec:(PPC-AWGN) Channel sims}
 
In this subsection, we apply the proposed approach for computing capacity and optimal input distribution using \ac{mine} method to the channel PPC-AWGN, introduced in Sec.  \ref{sec:PPC-AWGN Channel}.
%
%
%
When comparing the numerical results with the bounds in Th. \ref{thm:PPC} and Th. \ref{thm:PPC 2}, we observe that our approach 
either match or fill the gap between upper and lower bounds in each case.
This can be verified from in Tab. \ref{tab:PPC}, which reports the capacity approximation evaluated through the various \ac{ndt} networks, together with upper and lower bounds
%
%
in Th. \ref{thm:PPC} and Th. \ref{thm:PPC 2}. 

\begin{table*}
\centering
{
\small
\begin{tabular}{ |c|c|c|c|c|c| } 
 \hline
  SNR ($\rm dB$ ) & $2$ &$ 5$ & $10$ & $15$ & $20$ \\ 
 \hline
 \ac{mine} ($\nats$)  &$0.426\pm 0.003$&$0.510\pm 0.007$ &$0.910 \pm .006$& $1.260 \pm 0.035 $ & $1.710 \pm 0.004 $\\
 \hline
 \ac{smile} ($\nats$) & $0.474\pm 0.002 $&$0.570 \pm 0.005 $ & $0.940\pm 0.007$& $1.320\pm 0.01$ & $1.830\pm 0.011$\\ 
 \hline
 \ac{dine} ($\nats$) &$0.410\pm 0.082$& $0.540\pm 0.01$ & $0.930\pm 0.02$&  $1.280 \pm 0.03$ &$1.680\pm 0.03$\\
 \hline
 NWJ ($\nats$) &$0.44\pm .0002$&$0.56\pm .0004$&$0.911\pm 0.0005$&$1.25\pm 0.0055$&$1.72\pm 0.0075$\\
 \hline
 Chi-Square ($\nats$)& $0.42\pm 0.0101$&$0.55\pm 0.02$&$0.91\pm .02$&$1.241\pm 0.3$&$1.71\pm 0.05$\\
 \hline
 DV-Variant ($\nats$)& $0.465\pm 0.001$& $0.59\pm 0.003$&$0.91\pm 0.008$&$1.31\pm 0.01$&$1.77\pm .03$ \\
 \hline
Lower bound ($\nats$) & $0.410$ & $0.530$ & $0.910$ & $1.230$ & $1.700$ \\
\hline
Upper bound ($\nats$) & $0.474$ & $0.620$ & $0.927$ & $1.330$ & $1.780$ \\
\hline
\end{tabular}
}
\caption{Numerically optimized  capacity through the various \ac{ndt} networks for the PPC-AWGN channel, together with upper and lower bounds in Th. \ref{thm:PPC} and Th. \ref{thm:PPC 2}. As in \cite{kramer},  simulations consider the case  $A=\sqrt{P} \ {\rm dB}$.
}
\label{tab:PPC} 
\end{table*}

\begin{figure}[t!]
    
                

\begin{tikzpicture}
    \definecolor{mycolor1}{rgb}{0.00000,0.44706,0.74118}%
    \definecolor{mycolor2}{rgb}{0.63529,0.07843,0.18431}%
    \definecolor{mycolor3}{rgb}{0.00000,0.49804,0.00000}%
    
    \begin{groupplot}[
        group style={
            group name=my plots,
            group size=3 by 1,
            xlabels at=edge bottom,
            xticklabels at=edge bottom,
            vertical sep=15pt,
            horizontal sep=2pt
        },
        height=4cm,
        width=7.25cm,
        yticklabels={,,}
        xmajorgrids=true,
     ]
]


\nextgroupplot[title = {case $A=\sqrt{P}=2 {\rm dB}$}
,xmajorgrids=true,ymajorgrids=true,
]

 \addplot [ybar interval, fill=mycolor1!120,draw=mycolor1]
 table[col sep=comma]{./Data/data-awgn-peak=1,avg=2db.txt};
 
\nextgroupplot[title = {case $A=\sqrt{P}=10 {\rm dB}$},
,xmajorgrids=true,ymajorgrids=true,
]
%

\addplot [ybar interval,fill=mycolor1!120,draw=mycolor1]
    table[col sep=comma]{./Data/data-awgn-peak=5,avg=10db.txt};

\nextgroupplot[title = {case $A=\sqrt{P}=20 {\rm dB}$}
,xmajorgrids=true,ymajorgrids=true,
]
%

\addplot [ybar interval,fill=mycolor1!120,draw=mycolor1]
    table[col sep=comma]{./Data/data-awgn-peak=10,avg=20db.txt};

\end{groupplot}
\end{tikzpicture}
    
\label{fig:ppc}
\caption{Computed optimal input distribution by NIT for the PPC-AWGN channel discussd in Sec.  \ref{sec:(PPC-AWGN) Channel sims} with $A=\sqrt{P}\in \{2,10,20\}  {\rm dB} $.
%
}
\label{in_peak}
\end{figure}
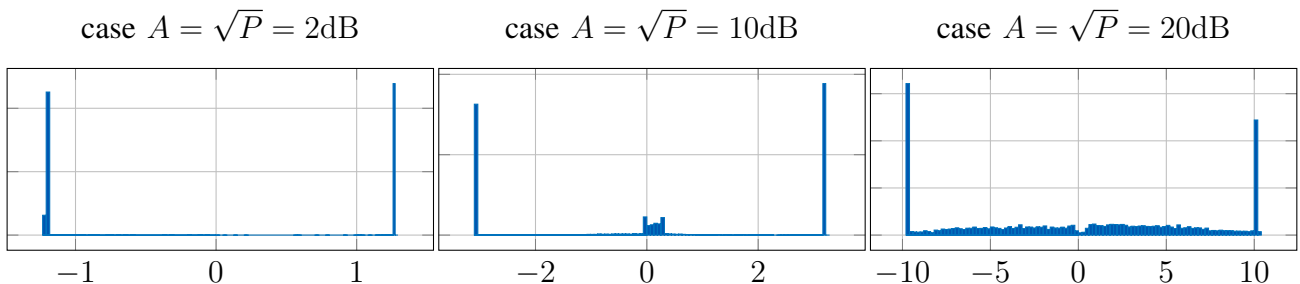


%

Based on the numerical evaluations, we conclude that 
using \ac{nmie} methods that estimate the MI directly can result in more accurate estimates of the channel capacity and the optimal input distribution. While more investigation is needed to gain a deeper understanding of this observation, we suspect that this is because the error terms in entropy-based estimations 
compound through the learning process, resulting in less accurate estimates of MI and, consequently, capacity. Moreover, entropy-based estimators tend to be less robust to the random initialization of the networks.


  

\subsection{The Poisson Channel}
\label{sec:The Poisson Channel}

Next, let us investigate the capacity of the Poisson channel in Sec. \ref{sec:Poisson Channel} to further understand the role of discrete initialization of proposed deep estimation approach as detailed in Algorithm \ref{alg:NCE_discrete}. As we mentioned earlier, the motivation for the use of discrete initialization is that for the Poisson channel, where the output of the channel and the optimal distribution are discrete, discrete initialization distributions achieve higher information rates compared to using normal distribution as initialization distributions which are closer to the capacity. This has been illustrated in the simulation summarized in the Table \ref{tab:Pois} for the Poisson channel.
%
%

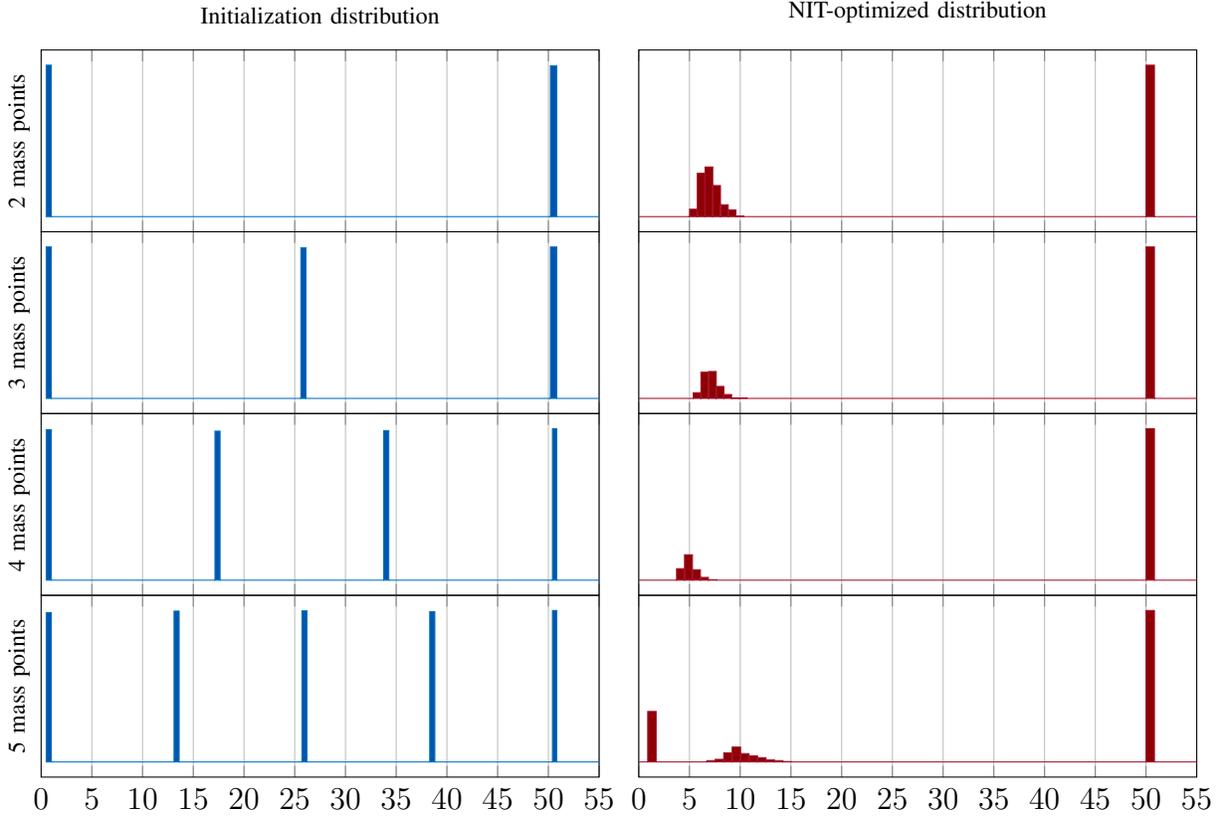
\begin{figure}
    \centering
    \begin{tikzpicture}
    \definecolor{mycolor1}{rgb}{0.00000,0.44706,0.74118}%
    \definecolor{mycolor2}{rgb}{0.63529,0.07843,0.18431}%
    \definecolor{mycolor3}{rgb}{0.00000,0.49804,0.00000}%
    
    \begin{groupplot}[
        group style={
                group name=my plots,
                group size=2 by 4,
                xlabels at=edge bottom,
                xticklabels at=edge bottom,
                vertical sep= 0pt,
                horizontal sep=15pt
                },
        height=4cm,
        width=9cm,
        ytick=\empty,
        xmajorgrids=true,
        ymajorgrids=true,
     ]
]


\nextgroupplot[
            legend style={legend columns=4},%
            ylabel={\footnotesize 2 mass points},%
            title style={text depth = 0pt},  
            title={\footnotesize Initialization distribution},
,xmin=0,xmax=55,xtick={0,5,...,55},xmajorgrids=true,ymajorgrids=true,]

\addplot [ybar interval,mark=no, fill=mycolor1!120,draw=mycolor1]
 table[col sep=comma]{./data_m/init-poisson-2p.txt};

\nextgroupplot[
title={\footnotesize NIT-optimized distribution},
,xmin=0,xmax=55,xtick={0,5,...,55},xmajorgrids=true,ymajorgrids=true,]

\addplot [ybar interval,mark=no,fill=mycolor2!120,draw=mycolor2]
    table[col sep=comma]{./data_m/data-poisson-out-2points.txt};


\nextgroupplot[
ylabel={\footnotesize 3 mass points},%
,xmin=0,xmax=55,xtick={0,5,...,55},xmajorgrids=true,ymajorgrids=true,
]

\addplot [ybar interval,mark=no,bar width=2pt, fill=mycolor1!120,draw=mycolor1]
 table[col sep=comma]{./data_m/init-poisson-3p.txt};

\nextgroupplot[
,xmin=0,xmax=55,xtick={0,5,...,55},xmajorgrids=true,ymajorgrids=true,]

\addplot [ybar interval,fill=mycolor2!120,draw=mycolor2]
    table[col sep=comma]{./data_m/data_poisson_out_3points_.txt};


\nextgroupplot[
ylabel={\footnotesize 4 mass points},%
,xmin=0,xmax=55,xtick={0,5,...,55},xmajorgrids=true,ymajorgrids=true,]

\addplot [ybar interval,mark=no,bar width=2pt, fill=mycolor1!120,draw=mycolor1]
 table[col sep=comma]{./data_m/init-poisson-4p.txt};

\nextgroupplot[
,xmin=0,xmax=55,xtick={0,5,...,55},xmajorgrids=true,ymajorgrids=true,]

\addplot [ybar interval,fill=mycolor2!120,draw=mycolor2]
    table[col sep=comma]{./data_m/data_poisson_out_4point_.txt};


\nextgroupplot[
ylabel={\footnotesize 5 mass points},%
,xmin=0,xmax=55,xtick={0,5,...,55},xmajorgrids=true,ymajorgrids=true,]

\addplot [ybar interval,mark=no,bar width=2pt, fill=mycolor1!120,draw=mycolor1]
 table[col sep=comma]{./data_m/init-poisson-5p.txt};

\nextgroupplot[
,xmin=0,xmax=55,xtick={0,5,...,55},xmajorgrids=true,ymajorgrids=true,]

\addplot [ybar interval,fill=mycolor2!120,draw=mycolor2]
    table[col sep=comma]{./data_m/data_poisson_out-5points-.txt};

\end{groupplot}

\end{tikzpicture}
    \caption{
    A representation of the input distribution optimization for the Poisson channel according to  Algorithm \ref{alg:NCE_discrete}. 
    %
    A representation of the input distribution optimization as in Algorithm \ref{alg:NCE_discrete} for the Poisson channel in Sec. \ref{sec:Poisson Channel} with $\Ecal=10 {\rm dB}$ and $A=50{\rm dB}$.
    In the left column we plot the initialization of the input distribution, on the right column the \ac{ndt}-optimized distribution.
    The different rows correspond to initializations with an increasing number of points in the initial optimal input distribution estimate. 
    }
    \label{init_m}
\end{figure}

For this model it is known \cite{shamai1}  that the capacity-achieving distributing is discrete and with a number of points that increases with the SNR.
In order to identify the cardinality of the optimal input distribution we proceed as in Algorithm \ref{alg:NCE_discrete}, that is by choosing a discrete initialization of this distribution with uniformly spaced mass points between zero and the peak power constraint (if considered).
An increasing number of mass points is considered in the initialization until the MI is no longer increasing.
%
%
The various stages of  Algorithm \ref{alg:NCE_discrete} are represented in Fig. \ref{init_m}.
%

%
%
%
%
%

Having established the capacity optimization approach, let us validate the usefulness of such approach  by considering the general Poisson channel in which both average and peak power constraints are present.

Since for the Poisson channel, a closed-form expression of the capacity and the optimal input distribution is not available, we consider those regimes in which tight lower and upper bounds on capacity are known in the literature.
More specifically it is known that \cite{shamai1},  in the case of zero dark current (that is $\lambda=0$) and only peak constraint,  the optimal input, is binary in the regime  $0\leq A \leq 3.3679$ and has larger cardinality when 
$A\geq 3.3679$.
%
Furthermore in \cite{chen1} it has been proved that, for the Poisson channel with both peak and average constraint, the optimal input always contain a mass point at zero.
In addition \cite{chen1} show that, under only peak power constraint, the optimal input distribution contains a mass point at the peak value $A$. 
In our simulations, we consider the case of a Poisson channel under average power constraint only.
In this regime,  \cite{shamai1} and \cite{chen1} conjecture that the optimal input distribution is exponential. 

\begin{figure}
    \centering
    \begin{tikzpicture}
    \definecolor{mycolor1}{rgb}{0.00000,0.44706,0.74118}%
    \definecolor{mycolor2}{rgb}{0.63529,0.07843,0.18431}%
    \definecolor{mycolor3}{rgb}{0.00000,0.49804,0.00000}%
    
    \begin{groupplot}[
        group style={
            group name=my plots,
            group size=2 by 1,
            xlabels at=edge bottom,
            xticklabels at=edge bottom,
            vertical sep=15pt,
            horizontal sep=20pt
        },
        height=4cm,
        width=9cm,
        yticklabels={,,}
        xmajorgrids=true,
     ]
]


\nextgroupplot[title = {case $A=\Ecal=3$}
,xmajorgrids=true,ymajorgrids=true,
]

\addplot [ybar interval, fill=mycolor1!120,draw=mycolor1
,opacity=0.5,] table[col sep=comma]{./Data/poisson_3db_binary_data.txt};

\addplot [ybar interval, fill=mycolor2!120,draw=mycolor2
,opacity=0.5,
] table [col sep=comma]{./Data/shpois3db.txt};

\nextgroupplot[title = {case $A=\Ecal=4$},
,xmajorgrids=true,ymajorgrids=true,
]


\addplot[ybar interval, fill=mycolor1!120,draw=mycolor1,
opacity=0.5,] 
table[col sep=comma]{./Data/poisson_4db_ternary_data.txt };

\addplot [ybar interval, fill=mycolor2!120,draw=mycolor2,
opacity=0.5,] 
table[col sep=comma]{./Data/pois_chen_4db.txt};

\end{groupplot}
\end{tikzpicture}

\caption{{ \fm Computed optimal input distribution by NIT for the Poisson channel presented in Sec.  \ref{sec:Poisson Channel} with $A=\Ecal \in \{3,4\}$. 
Note that the transition between two to three mass points in optimal input distribution occurs for $A=3.36$ \cite{shamai1}.
}
%
}
\label{fig:poisson}
\end{figure}
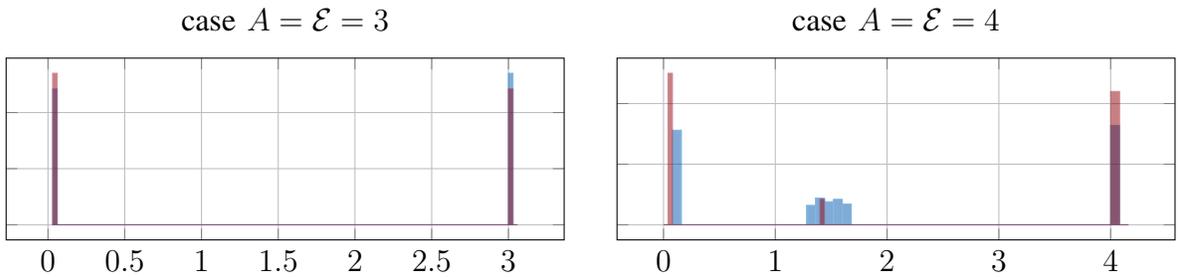

Is Fig. \ref{fig:poisson}, we plot the simulations results for  the Poisson channel  for $\Ecal=A \in \{3,4\}$. 
For the case $\Ecal=A=3$, the capacity values provided in 
\cite{shamai1} is $0.5944 \nats$ while the estimated capacity through MINE is $0.5852 \nats$.
Note that the proposed approach estimates an optimal input distribution with binary support. 
Also note that, according to \cite{6685986} the optimal input distribution always have a mass point at peak constraint.
For the case  $\Ecal=A=4$, instead, the optimal input distribution contains three mass points. This is correctly recovered by the proposed approach. 
For this scenario, the capacity estimated in   \cite{caocapacity11} is $0.91\nats$, while our approach yields  an estimated capacity equal to $0.82\nats$.

In each scenario, we compare the performance of MINE with the upper and lower bound in Th. \ref{th:poisson}, originally from  \cite{moser}. 

In Table \ref{tab:Pois} we provide a similar comparison of the estimated capacity for the Poisson channel with $A=100$ and $\Ecal \in \{ 5,10,15,20\}$.
Here we compare the MINE, the lower bound in Th. \ref{th:poisson} as well as the deterministic annealing (DA) approach from\cite{caocapacity11}.
Note that, in Table \ref{tab:Pois}, we also provide the performance of the \ac{ndt} in estimating capacity for both the Gaussian initialization as well as the discrete initialization devised in Algorithm \ref{alg:NCE_discrete}. 
It can be observed that, for this model, the performance of the discrete initialization far outperforms that of the Gaussian one.  



 \begin{table*}
 \centering
 {
\small
 \begin{tabular}{|c|c|c|c|c|c|} 
  \hline
   SNR ({\rm dB})   &$5$& $10$ & $15$ &$ 20$& Dark current \\ 
  \hline
   \ac{mine} (Discrete) ($\nats$)   &$1.08 \pm 0.0022 $&$1.44\pm 0.0031$ &$1.73\pm 0.0042 $& $1.74\pm 0.0050$ &$\lambda=0.0$\\ \hline
   \ac{mine} (Gaussian) ($\nats$) &$0.62\pm 0.0011$& $0.94\pm0.0015$ &$1.42\pm 0.0020 $& $1.46\pm 0.0037$ &$\lambda=0.0$\\
 \hline
  Chi-Square   &$0.422\pm 0.0033$& $0.7611\pm0.01$ & $1.2839\pm 0.022$ & $1.7687\pm0.041$ &$\lambda=0.0$\\ \hline
 
  DA ($\nats$)   &$0.95$&$1.39$& $1.78$ &$1.79$&$\lambda=0.0$\\ \hline
 
  Lower bound ($\nats$)   &$0.0$&$ 1.02$,&$1.55$&$1.55$&$\lambda=0.0$\\ \hline

 \ac{mine} (Discrete) ($\nats$)  &$0.554\pm 0.0021 $&$1.09\pm 0.0031 $ &$ 1.2\pm 0.0041$& $1.4\pm 0.0052$  &$\lambda=10.0$\\ \hline
\ac{mine} (Gaussian) ($\nats$) &$0.24\pm 0.0011$& $0.54\pm 0.0013$ &$0.86\pm 0.0021 $& $0.097\pm 0.0022$ &$\lambda=10.0$\\ \hline

 DA ($\nats$)  &$0.51$&$0.91$&$1.5$,&$1.51$&$\lambda=10.0$\\ \hline

 Lower bound ($\nats$)  &$0.0$&$0.0$&$1.23,$&$1.23$ &$\lambda=10.0$\\
 \hline
 \end{tabular}
 }
 \caption{
   Comparison of in the estimation of the capacity of the Poisson channel with $A=100$ and $\Ecal \in \{ 5,10,15,20\}
  {\rm dB}$ using MINE with discrete initialization, MINE with Gaussian initial distribution, the deterministic annealing (DA) approach from\cite{caocapacity11}, and the lower bound in Th. \ref{th:poisson}. MINE neural estimates are averaged over 10 trials.
 }
 \label{tab:Pois}
 \end{table*}

\begin{figure}
\centering
\begin{tikzpicture}
\definecolor{mycolor1}{rgb}{0.00000,0.44706,0.74118}%
\begin{axis}[name=plot,xmin=0,xmax=100,width=.5\textwidth,grid=major
,height=4cm,width=12cm,
ytick=\empty,
ymajorgrids=true,
]
\addplot [ybar interval, fill=mycolor1!120,draw=mycolor1] 
table[col sep=comma]{./Data/no_avg_10db.txt};
\end{axis}
\end{tikzpicture}
\caption{\ac{ndt}-optimized input distribution for the Poisson channel with only an average power constraint of $\Ecal=10 {\rm dB}$. }
\label{fig:only avg}
\end{figure}
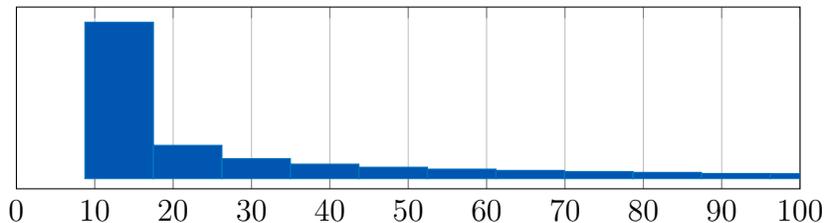

Finally, in Fig. \ref{fig:only avg}, we plot the 
Poisson channel with only average power constraint. 
In \cite{shamai1}, the authors conjecture that the optimal input distribution has unbounded support and takes shape of exponential distribution.
We notice that Fig. \ref{fig:only avg} indeed seems to validate this conjecture.
For this model,  the \ac{ndt}-estimated capacity is $1.88 \nats$, the lower bound in Th.  is $0.991 \nats$ while the asymptotic upper bound from \cite{moser} is $1.19 \nats$.


  \subsection{Multiple Access Channel (MAC)}
\label{sec:Multiple Access Channel (MAC)}
In this section we apply MINE method to compute the capacity of MAC introduced in Sec. \ref{sec:Multiple Access (MAC) Channel}.
To compute the sum capacity using MINE method we produce two sequence of i.i.d random variables and after concatenating them we feed them to \ac{ndt} neural networks. Then, each column of the output matrix is used as two-input distribution of the channel which satisfy the channel constraints. 
After producing the output of channel using these inputs from \ac{ndt} network, we use \ac{nmie} networks to estimate the mutual information and train the joint network of \ac{ndt} and {NMIE} based on the Algorithm \ref{alg:NCE}.
The corner point of capacity region are computed by the conditional mutual information $I(X_1;Y\mid X_2)$ and $I(X_2;Y\mid X_1)$.
To compute these points we simply decompose the mutual information as $I(X_1;Y\mid X_2)=I(X_1,X_2;Y)-I(X_2;Y)$ and use previous approach for each term to compute the conditional term.
%
For both the AWGN MAC and the OI MAC, we observe that \ac{mine} 
provides relatively accurate estimates of sum-capacity.
Furthermore, we observe that for the OI MAC, our approach provides estimates of achievable rates  which fall between the tightest inner and outer bounds for this model.

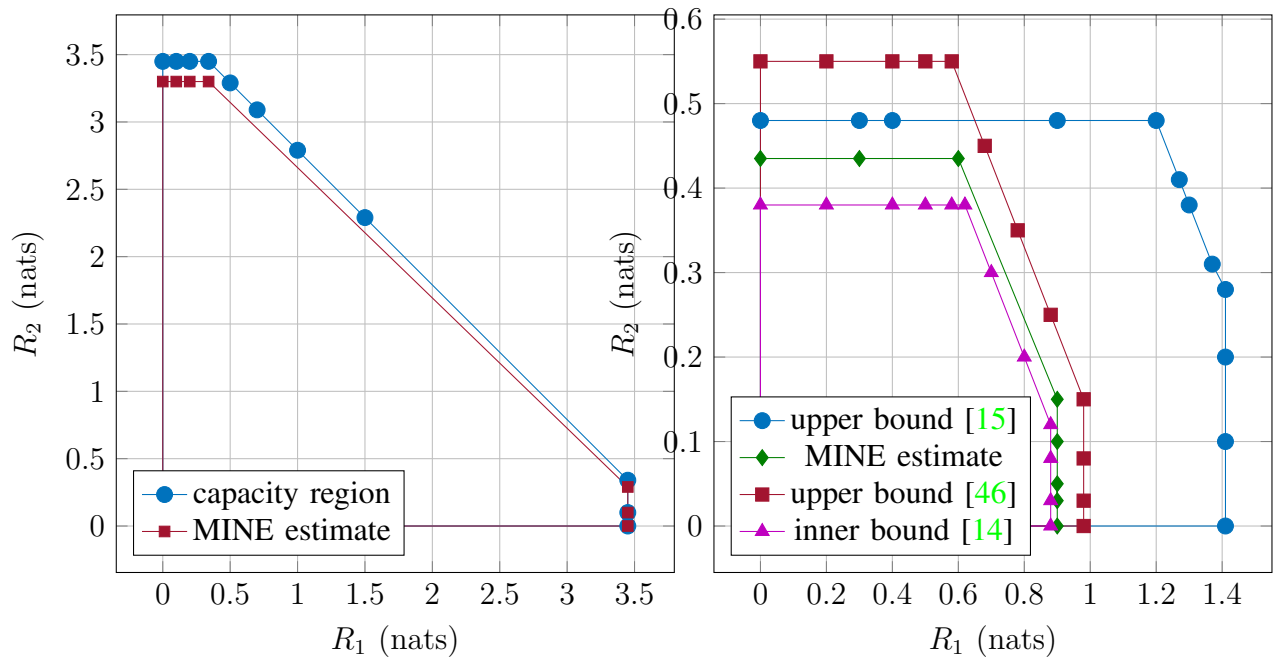
\begin{figure}
    \centering
    \begin{tikzpicture}
    \definecolor{mycolor1}{rgb}{0.00000,0.44706,0.74118}%
    \definecolor{mycolor2}{rgb}{0.63529,0.07843,0.18431}%
    \definecolor{mycolor3}{rgb}{0.00000,0.49804,0.00000}%
   \definecolor{mycolor4}{rgb}{0.00000,0.44700,0.74100} %
\definecolor{mycolor5}{rgb}{0.74902,0.00000,0.74902} %
 
    \begin{groupplot}[
        group style={
            group name=my plots,
            group size=2 by 1,
            xlabels at=edge bottom,
            xticklabels at=edge bottom,
            vertical sep=15pt,
            horizontal sep=15pt
        },
        height=9 cm,
        width=9 cm,
        xmajorgrids=true,
        ymajorgrids=true,
     ]
]


\nextgroupplot[
 xlabel={$R_1$ $(\nats)$},ylabel={$R_2$ $(\nats)$}
 ,legend entries={capacity region,\ac{mine} estimate},legend pos={south west}]
 ]
%

 \addplot  [draw=mycolor1,,mark size=3pt,mark=*,mark options={solid,draw=mycolor1,fill=mycolor1}]
 table[col sep=space]{./Data/doc-mac-e1.txt};
 
  \addplot [draw=mycolor2,mark size=2pt,mark=square*,mark options={solid,draw=mycolor2,fill=mycolor2}]
 table[col sep=space]{./Data/doc-mac-e2.txt};

\nextgroupplot[
 xlabel={$R_1$ $(\nats)$},ylabel={$R_2$ $(\nats)$},
,legend entries={upper bound \cite{oimac2},\ac{mine} estimate,upper bound \cite{kellips},inner bound \cite{oimac1}},legend pos={south west}]

\addplot 
[draw=mycolor1,,mark size=3pt,mark=*,mark options={solid,draw=mycolor1,fill=mycolor1}]
table [col sep=space]{./Data/doc-mac-opt.txt};
\addplot 
[draw=mycolor3,,mark size=3 pt,mark=diamond*,mark options={solid,draw=mycolor3,fill=mycolor3}]
table [col sep=space]{./Data/doc-mac-est.txt};

\addplot 
[draw=mycolor2,,mark size=2.5 pt,mark=square*,mark options={solid,draw=mycolor2,fill=mycolor2}]
table {./Data/doc-mac-opt1.txt};

\addplot 
[draw=mycolor5,mark size=3 pt,mark=triangle*,mark options={solid,draw=mycolor5,fill=mycolor5}]
table [col sep=space]{./Data/doc-mac-opt-in.txt};

\end{groupplot}

\end{tikzpicture}
 \caption{ Left:  A comparison of the capacity  region for AWGN MAC channel discuss in Sec. \ref{sec:AWGN MAC channel}  with the \ac{mine} estimate. 
 The choice of constraints is $P_1=20 {\rm dD}$ and $P_2=20{\rm dB}$.
  -- Right:  A comparison of inner and outer bounds of the capacity region of th OI MAC channel model in Sec. \ref{sec:Optical Intensity (OI) MAC channel} with the \ac{mine} estimate. For these evalutaions, we have chosen ${\mathbb{E}[X_1]}/{A_1}={\mathbb{E}[X_2]}/{A_2}=0.2$, $A_1=10 \rm{dB}$, $A_2=5\rm {dB}$, and $\sigma^2=1$.
 }
    \label{fig:my_label}
\end{figure}

 \section{Conclusion}
\label{sec:Conclusion}

{\ste
In this paper, we make a first attempt in establishing a benchmark for neural capacity estimation. 
We consider the problem of capacity estimation on a number of point-to-point channel, as well as multiple access channel (MAC). 
}
More precisely, we consider point-to-point channels (i) with a known capacity --AWGN channel 
as well as  (ii) channels with unknown capacity and continuous inputs – i.e., optical intensity and
peak-power constrained AWGN channel (iii) channels with unknown capacity and discrete number of mass
points – i.e., Poisson channel. 
We also extend the analysis to the AWGN MAC --case (i) above-- and the optical intensity MAC -case (ii). 
{\ste
For the models above, we study the performance of mutual information neural
estimator (MINE), smoothed mutual information lower-bound estimator (SMILE), and directed information
neural estimator (DINE). and provide insights on the performance of other methods as well.
}
%
%
Our numerical evaluations show that all methods perform rather well at low SNR, while only some methods perform well at high SNR.
In particular,  MINE emerges as the most versatile method across the different channel model.
From our numerous numerical experiments, we observe that the network initialization plays a crucial role attaining accurate  capacity estimation. 
For this reason, we propose a specific initialization strategy for channel which the optimal input distribution has a discrete number of masses, such as the Poisson channel. 
%
%
%
%
Overall, we notice that, although very promising, accurate neural capacity estimation 
%
crucially depends an the ability of the underlying network to approximate functions with a large support: future research will focus on the development of such DNN. 

\bibliographystyle{IEEEtran}
\bibliography{IEEEabrv,main}
\section{Appendix}
\normalsize
\quad\quad\quad\quad\quad\quad\quad\quad\quad\quad\quad\textcolor{blue}{\textbf{Consistency proof for $\chi^2$ bound :}}\newline
 \textcolor{blue}{We first discuss that expectation in the bound can be replaced by emperical averages over batches of data in three steps each for consistent estimation of each component $(\mathbb{E}_{P=(X,Y)}[T(X,Y)] - \mathbb{E}_{Q=(X\times Y)}\log[e^{T(X^\prime,Y^\prime)}]$,\newline$\chi^2(X||X^{'},\chi^2(Y||Y^{'})$. Then we prove our main consistency result for the$\chi^2$ estimator.
For the estimation of the first term we use expectations over mini-batches of data, by using law of large numbers for any $\epsilon$ and sufficiently large $n$ we have:
 \begin{flalign}
 |\mathbb{E}_{P}[T(X,Y)] -\frac{1}{n}\sum_{0}^{n-1} T^{*}(x_i,y_i)|\leq \epsilon\label{pf0}
 \end{flalign}
 where the optimal function $T^{*}$ is $T^{*}=\log \frac{dP}{dQ}+k$ where $k$ is a constant defined such that $dP$ is relate to Gibbs density by $dP=\frac{1}{k}e^{T^{*}}dQ$. For the second term we have:  \begin{flalign}
  \mid \mathbb\log({E}_{Q}[e^{T^(X^{'},Y^{'})}])-\log(\frac{1}{n}\sum_{0}^{n-1}e^ {T^\ast{}^{(x_{i}^{'},y_{i}^{'})}})\mid\leq \epsilon
   \end{flalign}
  to see that above inequality is true, it sufficies to note that $x_{i}^{'},y_{i}^{'}$ and their transformation through exponential function are iid random variables, and invoke AEP theorem\cite{cover}. by choosing a large $n$, $\epsilon$ could be made arbitrary small. Now we turn to  $\chi^{2}_{\mathsf{UP}}(P||Q)$ terms.We provide the argument for one term, and similar arguments hold for the last term. We have
 \begin{flalign}
 \chi^2(X||X^{'})=\mathbb{E_{X}}\left[T(X)\right] - \log \lb \mathbb{E}_{X^{'}}\left[T^{2}(x,x_i)\right] \rb-1
 \end{flalign}
 The estimation of first term considered in \eqref{pf0}, the estimation of second term using averages over mini-batches is also evident from weak law of large numbers and noting that $x_i,x_{i}^{'}$ are i.i.d.}\newline
 \textcolor{blue}{Now we turn to our main consistency result. For the first term in \eqref{eq:chisqrbound} the consistency of estimators using neural networks as universal approximators have been established in \cite{belghazi2018mine}\cite{hornik1989}. Here we provide consistency result for the variational form $\chi^2$ and show that it can provide consistent estimates when feed-forward neural network are used for its representation.  In fact we show the following:}
 \textcolor{blue}{\begin{theorem}
 There exist a family of neural network functions $T_{\theta}$ with parameters $\theta$ such that ($P,Q$ are probability measures as in previous theorems)
 \begin{flalign}
 |\chi^2(P||Q)-\hat{\chi}^2(P||Q)|\leq \epsilon
 \end{flalign}
 where we have : $\hat{\chi}^2(P||Q)=\sup_{T \in \mathcal{T}} \mathbb{E}_{P}[T(X)] - \log \lb \mathbb{E}_Q\left[{T^{2}(X)}\right] \rb-1$
 \end{theorem}}
 \textcolor{blue}{\textit{Proof.}\newline
 It is easy to show that (by using definition of the $\chi^2 (P||Q)=\mathbb{E}_{Q}\Bigg[(\frac{dP}{dQ}-1)^2\Bigg]$ and its variational form) the optimal function in \eqref{mainchi} is $T^{*}=\frac{dP}{dQ}$. Moreover, we can see we have:
 \begin{equation}\label{pf1}
  T^{*}=\frac{dP}{dQ}
 \end{equation}
 by construction we conclude
 \begin{equation}
     \chi^2(P||Q)=\mathbb{E}_{Q}\big[(\frac{dP}{dQ})^2-1\big]=\mathbb{E}_{Q}\big[(T^{*}(X))^2-1\big]
 \end{equation}}
 \textcolor{blue}{for an arbitrary function $T$ the error between approximated $\chi^2$ using this function and when optimal function is used is 
 \begin{flalign}
 &\chi^2(P||Q)-\chi^2(T)=\mathbb{E}_{Q}\big[(T^{*}(X))^2-1\big]+\mathbb{E}_{Q}(\big[(T(X))^2+1)-2\mathbb{E}_{P}\big[T(X)\big]\big]\nonumber\\
 &=\mathbb{E}_{Q}\big[(T^{*}(X))^2+T(X)^2\big]-2\mathbb{E}_{P}\big[T(X)\big]\label{pf2}\\
 &=\mathbb{E}_{Q}\big[(T^{*}(X))^2+T(X)^2-2T^{*}(X)T(X)\big]
 =\mathbb{E}_{Q}\big[(T^{*}(X))-T(X))^2\big]\label{pf4}
 \end{flalign}
 where to conclude \eqref{pf2} we have used $\mathbb{E}_{Q}\big[T^{*}(X)T(X)\big]=\mathbb{E}_{P}\big[T(X)\big]$, which can be derived from\eqref{pf1} by multiplying each side to $T(X)$. Now invoke the the universal approximation of neural network theorem\cite{hornik1989}. By this Theorem if the $T_{\theta}\leq M$ then we have:
 \begin{flalign}
 \mid \mathbb{E}_{P}\big[T^{*}-T_{\theta}\big]\mid\leq \frac{\eta}{2}\quad, \quad\quad \mid \mathbb{E}_{Q}\big[T^{*}-T_{\theta}\big]\mid\leq \frac{\eta}{2}e^{-M}
 \end{flalign}Thus, using the last line\eqref{pf4}, the error for the estimator is :
 \begin{flalign}
 |\chi^2(P||Q)-\chi^2(T)|= \frac{\eta^2}{4}e^{-2M}=\epsilon^{'}\leq \epsilon
 \end{flalign}This completes our proof.\newline Now it is trivial task to see that taking $\log$ and applying affine functions to this quantity preserves its consistent estimation and hence $\chi^{2}_{\mathsf{UP}}(P||Q)$.}
\normalsize

\end{document}